\documentclass[sigconf, noacm]{acmart}

\usepackage{color,soul}
\usepackage{caption}
\usepackage{float}
\usepackage[normalem]{ulem}

\usepackage{fancybox} 
\newcommand{\promptbox}[1]{
  \begin{center} 
    \doublebox{%
      \begin{minipage}{.95\columnwidth} 
        \vspace{5pt} 
        #1
        \vspace{5pt} 
      \end{minipage}%
    }
  \end{center}
}

\AtBeginDocument{%
  \providecommand\BibTeX{{%
    \normalfont B\kern-0.5em{\scshape i\kern-0.25em b}\kern-0.8em\TeX}}}

\setcopyright{none}

\acmDOI{}

\acmISBN{}

\newcommand{\revise}[1]{\textcolor{black}{#1}}

\newcommand{\rev}[1]{\textcolor{black}{#1}}

\begin{document}


\title{Why am I seeing this: Democratizing End User Auditing for Online Content Recommendations}

\author{Chaoran Chen}
\email{cchen25@nd.edu}
\affiliation{%
  \institution{University of Notre Dame}
  \city{Notre Dame}
  \state{Indiana}
  \country{USA}
}

\author{Leyang Li}
\email{lli27@nd.edu}
\affiliation{%
  \institution{University of Notre Dame}
  \city{Notre Dame}
  \state{Indiana}
  \country{USA}
}

\author{Luke Cao}
\email{lcao3@nd.edu}
\affiliation{%
  \institution{University of Notre Dame}
  \city{Notre Dame}
  \state{Indiana}
  \country{USA}
}

\author{Yanfang Ye}
\affiliation{%
  \institution{University of Notre Dame}
  \city{Notre Dame}
  \state{Indiana}
  \country{USA}
}

\author{Tianshi Li}
\email{tia.li@northeastern.edu}
\affiliation{%
  \institution{Northeastern University}
  \city{Boston}
  \state{Massachusetts}
  \country{USA}
}

\author{Yaxing Yao}
\affiliation{%
  \institution{Virginia Tech}
  \city{Blacksburg}
  \state{Virginia}
  \country{USA}
}

\author{Toby Jia-Jun Li}
\affiliation{%
  \institution{University of Notre Dame}
  \city{Notre Dame}
  \state{Indiana}
  \country{USA}
}

\renewcommand{\shortauthors}{Chen, et al.}

\ccsdesc[500]{Security and privacy~Human and societal aspects of security and privacy}
\ccsdesc[300]{Human-centered computing~Human computer interaction (HCI)}

\keywords{end-user auditing, LLM generated persona, privacy awareness}

\begin{abstract}
Personalized recommendation systems tailor content based on user attributes, which are either provided or inferred from private data. Research suggests that users often hypothesize about reasons behind contents they encounter (e.g., ``I see this jewelry ad because I am a woman''), but they lack the means to confirm these hypotheses due to the opaqueness of these systems. This hinders informed decision-making about privacy and system use and contributes to the lack of algorithmic accountability. To address these challenges, we introduce a new interactive sandbox approach. This approach creates sets of synthetic user personas and corresponding personal data that embody realistic variations in personal attributes, allowing users to test their hypotheses by observing how a website's algorithms respond to these personas. We tested the sandbox in the context of targeted advertisement. Our user study demonstrates its usability, usefulness, and effectiveness in empowering end-user auditing in a case study of targeting ads.

\end{abstract}
\maketitle



\section{Introduction}
Personalized recommendation systems customize content using personal profiles, catering to individual user interests~\cite{adomavicius2005toward, aggarwal2016recommender}. While these systems provide convenience, they often leave users feeling disempowered due to a lack of transparency in these algorithmic systems, making it challenging to understand the causal relationships between their personal data characteristics and the system's decisions~\cite{10.1145/3610209}. Prior research shows that users often have specific hypotheses about which personal data elements are used to target them in personalized recommendations~\cite{lee2023and}. For instance, some believe their browsing history influences the ads they see, particularly when ads relate to previously visited websites, while others speculate that factors such as age or location are more significant. These beliefs, whether accurate or not, shape users' mental models and influence their decision-making processes~\cite{yao2017folk, lee2023and}.

Instead of relying solely on passive transparency measures, such as system behavior disclosures by first parties~\cite{ehsan2021expanding, coglianese2019transparency} or audits by third parties~\cite{hussein2020measuring,haroon2023auditing}, an end-user auditing approach can empower users to actively investigate how their data characteristics affect the recommendations they receive~\cite{devos2022toward, 10.1145/3555625}. This proactive engagement in auditing and verifying how personal characteristics influence content recommendations can enhance users' privacy awareness and strengthen algorithmic accountability.

\begin{figure*}[ht]
    \centering
    \includegraphics[width=0.8\linewidth]{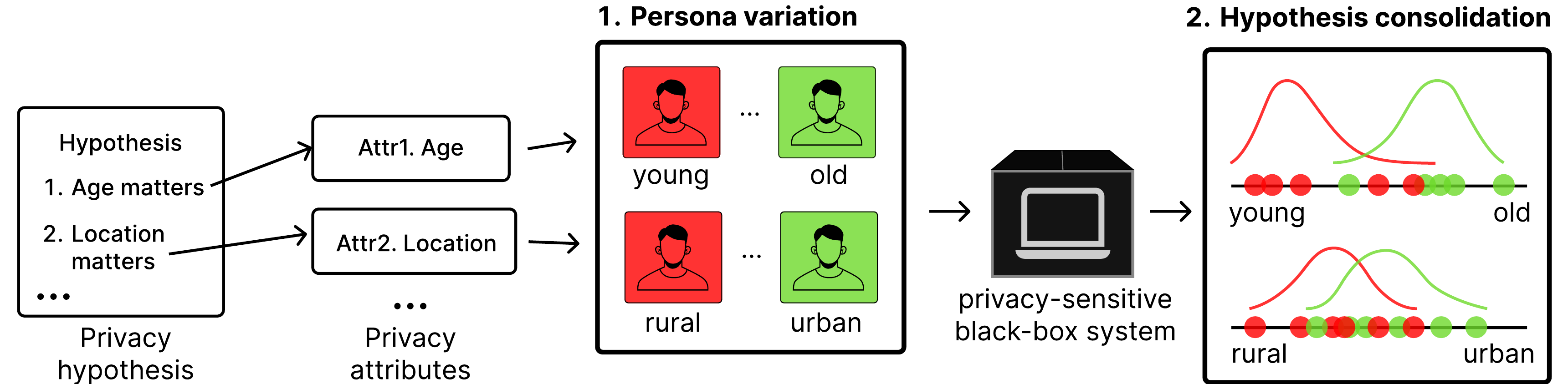}
    \caption{An overview of our privacy auditing sandbox approach}
    \Description{This figure provides an overview of a privacy auditing sandbox approach. The process begins with formulating privacy hypotheses, such as "age matters" or "location matters," based on privacy attributes (e.g., age, location). These attributes are used to create varied personas, such as young vs. old or rural vs. urban, which are then tested against a privacy-sensitive black-box system. The results of these tests are analyzed in the hypothesis consolidation phase, where differences in outcomes for each persona variation are visualized, helping to confirm or refute the initial privacy hypotheses.}
    \label{fig:overall_approach}
\end{figure*}

However, empowering end-users to audit online content recommendations presents significant challenges.

First of all, most current approaches do not align well with user needs.  
Traditional audits are primarily designed to meet regulatory requirements~\cite{courville2003auditing}, which do not necessarily clarify the relationship between users' data and system outcomes. Although privacy-enhancing tools such as anti-tracking measures~\cite{bubukayr2022web} and data encryption~\cite{deepika2020review} provide some level of data protection, these passive measures are insufficient to verify user privacy hypotheses, improve their privacy knowledge, or build trust in the system.
Additionally, the existing power imbalance between users and organizations further complicates matters, as privacy auditing practices often benefit organizations more than the individuals whose data is being audited~\cite{urman2024right}. 
Moreover, existing approaches often overlook users' agency in the auditing process, resulting in users' lack of engagement. Users typically have no active role in these processes and must passively accept the outcomes of audits conducted by technology companies or third-party agencies.
The lengthy duration of these audits also often renders their results outdated~\cite{kirkpatrick2024}, diminishing their usefulness for users who need real-time privacy validation. These third-party audits usually focus only on the most popular websites and hypotheses~\cite{sandvig2014auditing}, neglecting the diverse and specific auditing needs of many users, referred to as the ``long tail'' of user needs. Consequently, current privacy audits fail to empower users or provide them with timely, actionable insights, highlighting a clear need for more agile and user-centric auditing solutions.

Motivated by the aforementioned challenges and limitations of current end-user auditing, we introduce a novel user-centric approach called the Privacy Auditing Sandbox. The sandbox enables users to explore and validate their hypotheses regarding the causal relationships between their personal characteristics and the recommendations they receive online. The key novelty of our work is to introduce a new interactive sandbox approach that empowers non-technical users to generate evidence for or against their privacy hypotheses by directly interacting with recommendation systems in a controlled, risk-free environment. As shown in Fig.~\ref{fig:overall_approach}, this system includes two key features: persona variation and hypothesis consolidation.
\begin{itemize}
    \item \textbf{Persona variation}: 
    A major challenge in supporting end-user audits is enabling users to reason about how specific personal attributes influence system outputs. To address this, we introduce persona variation, a technique for systematically isolating and testing individual privacy attributes. We build on prior work~\cite{chen2023empathy} that used large language models (LLMs) to generate fictional personas enriched with demographic and longitudinal data. Drawing on variation theory~\cite{marton2014necessary}, our method extends this by creating multiple versions of a base persona, each mainly differing by one key attribute (e.g., age or location). This controlled variation allows users to compare system responses across personas and infer whether changes in attributes result in different recommendations.
    \item \textbf{Hypothesis consolidation}: 
    Even when users can observe differences in recommendations across persona variations, it can be difficult to interpret or summarize these changes, especially when outputs are unstructured (e.g., images, text). To address this, we designed hypothesis consolidation to help users make sense of observed patterns and draw evidence-based conclusions. After gathering data on how the system reacts to different personas, the system employs LLMs to map these outputs onto corresponding privacy attribute axes. For example, if exploring age, LLMs categorize output to show which age group each recommendation targets. This method extends to other attributes, allowing users to see how recommendations might change based on urban versus rural settings or across different income levels.  By visualizing how outputs shift across variations, users can more easily confirm or refute their initial hypotheses about which personal characteristics influence recommendations.
\end{itemize}

To evaluate the Privacy Auditing Sandbox, we conducted a case study in the context of online advertising, investigating how different privacy attributes influence targeted ad content. We first performed a technical evaluation to address core research questions related to persona generation quality, ad identification accuracy, ad rating stability, and the impact of persona substitution on advertisements. For persona generation, expert reviews highlighted strong consistency and credibility, although minor hallucinations and stereotyping were noted. For ad identification, a curated dataset of 205 websites and 805 advertisements demonstrated high accuracy (96.52\%) and precision, with minor errors stemming from dynamic or non-standard ad formats. LLM-based ad rating stability was confirmed with low variability (standard deviation = 0.87; coefficient of variation = 3.00\%) across privacy attributes. Lastly, persona substitution experiments revealed that distinctly different personas received significantly varied ads, while similar personas received consistent ads, underscoring the influence of persona attributes on ad content.

To evaluate our system's usability, we conducted a user study ($N=15$) and asked the participants to use pre-defined personas to test their privacy hypotheses and create new personas to explore further. We then assessed the usability with the the System Usability Scale (SUS) questionnaire. The results show that participants successfully validated their privacy hypotheses, identifying clear correlations between persona attributes and the thematic and visual aspects of the ads. Participants also reported a positive user experience and recognized the system’s value in enhancing their understanding of how personal characteristics affect ad targeting.

This approach has broader applications beyond targeted advertising, offering the potential for auditing the accountability of other privacy-sensitive systems. It is particularly useful in settings where generated persona data can be seamlessly integrated into systems, allowing users to repeatedly prompt system responses in real time at minimal cost. \revise{We anticipated that the tool can enable users to (1) discover unexpected uses of their personal data and (2) validate concerns related to algorithmic practices in content personalization, such as biases in personalized advertising.} This study represents an initial step towards enabling robust end-user auditing of online content recommendations and enhancing end-user privacy awareness. 

This paper makes the following contributions: 
\begin{itemize}

    \item Introduction of a novel interactive approach for non-technical end users to audit online recommendations and explore and verify their hypothesis about causal relationships between personal privacy attributes and system outcomes in a risk-free environment through LLM-generated personas for end-user auditing. \revise{We make a unique technical contribution by utilizing state-of-the-art methods for generating personas enriched with demographic data and longitudinal personal data~\cite{chen2023empathy}.}
    \item Validation of the approach through a proof-of-concept prototype and a case study. The study results reveal patterns in how users employ our method to validate their privacy hypotheses in the context of targeted advertising.
    \item Discussion of the implications of adopting this approach to enhance user engagement in privacy auditing and improve their privacy awareness. 
\end{itemize}

\section{Related work}
\label{sec: related work}



\revise{
We reviewed prior work in explainable recommendation systems, algorithm auditing, and counterfactual reasoning to address the challenge of enhancing user understanding and accountability in algorithmic decision-making. Explainable recommendation systems aim to uncover causal relationships between recommended content and user characteristics. However, their technical complexity and limited applicability in real-world contexts often hinder end users from fully grasping these mechanisms. Recent advances in algorithm auditing have incorporated users into the auditing process, enabling them to explore and evaluate the causal connections between privacy attributes and recommended content, thereby increasing privacy awareness and strengthening algorithmic decision-making. Counterfactual reasoning further supports these efforts by illustrating the links between private data and online content recommendations through comparison and contrast. Lastly, as we used online behavioral advertising (OBA) as a case study, we review and contrast common static methods for improving OBA transparency, underscoring the novelty of our proposed interactive Privacy Auditing Sandbox system: it allows end users to learn through exploration, actively verify privacy hypotheses, and observe correlations between privacy attributes and system outcomes.
}

\subsection{Explainable Recommendation}
Personalized recommendation systems use vast amounts of user data to customize content based on the preferences and behaviors of individual users. To capture the causal relationship between the recommended content and user characteristics, a popular solution is ``explainable recommendation''~\cite{zhang2014explicit}. This approach aims to not only offer recommendation outcomes to users but also provide explanations for why specific items are recommended. It has been widely applied to diverse recommendation systems, such as those in e-commerce~\cite{gao2022survey}, points of interest~\cite{baral2017pers}, social networks~\cite{guy2011social}, and multimedia~\cite{katarya2017effective}. 

The explanations of recommendation models have different target audiences (e.g., users, model designers, and service providers) and various objectives (e.g., improving informed decision-making and enhancing profitability). Therefore, different target audiences and objectives lead to diverse evaluation perspectives. The most related criteria for understanding the causal relationship between the recommended content and user data is transparency, which evaluates whether the explanations can reveal the internal working mechanism of the recommendation models~\cite{tai2021user, sonboli2021fairness, li2021attribute}. For example, Jiang et al. proposed RCENR, a Reinforced and Contrastive heterogeneous Network Reasoning model for explainable news recommendation~\cite{jiang2023rcenr}, which generates multi-hop reasoning paths to connect users with candidate items through intermediate nodes (e.g., previous clicked news and the attributes of the news). Zhang et al.~\cite{zhang2022neuro} presented NS-ICF, a neuro-symbolic interpretable collaborative filtering that learns logical rules for recommendation systems. 

However, the objectives of the above work focus on assisting model designers in knowing more about how the recommendations are generated and better debugging the system. End users often find it hard to understand these model outputs because of technical complexities and difficulties in applying these explainable models in real life. They struggle to see how their private data leads to specific recommendations. Compared with such methods assisting model designers, our system is designed to empower end users to understand and experiment with recommendation systems, enabling them to observe how different privacy attributes of generated personas can influence recommendation items.

\subsection{Algorithm Auditing for Online Content Recommendation}

Algorithm audits are effective tools for analyzing black-box systems by systematically querying them with inputs and observing the outputs~\cite{metaxa2021auditing}. These audits have been applied across various domains, such as employment~\cite{speicher2018potential}, healthcare~\cite{obermeyer2019dissecting}, and consumer markets~\cite{ali2019discrimination}, to promote fairness and transparency~\cite{metaxa2021auditing}. 

Traditionally, auditors have excluded real users to maintain control and strengthen technical audits~\cite{datta2014automated, scharowski2023certification}. The typical methodology involves repeatedly querying an algorithm with inputs and observing the outputs to infer the system's internal workings. Recent studies incorporate end users into the auditing process. There are two main types of input data for algorithm audits: one relies on crowdsourcing real users to conduct audits, while the other involves simulating users by generating data with specific attributes. For crowdsourcing, Lam et al.~\cite{10.1145/3610209} introduced sociotechnical audits, demonstrating how real users can be actively involved in algorithm audits by systematically collecting their input and behavioral data to reveal algorithmic biases. 
Although using crowdsourcing ensures that the audit involves real users, this method requires a significant amount of time and resources to engage many participants. To improve auditing efficiency while still using user-related data, simulating user data for audits has been developed.
For example, Sunlight~\cite{lecuyer2015sunlight} introduces a scalable system that uses statistical methods to detect the causes of targeting phenomena, such as personalized ads in Gmail, by simulating users with different attributes.

Compared to Sunlight, our approach is not limited to Gmail but is applicable to ads across various websites. We use large language models (LLMs), which allow for more detailed simulations of users' longitudinal data, such as their weekly schedule and browsing history. Additionally, we build on the persona generation technique proposed by Chen et al.~\cite{chen2023empathy}, generating not just a single persona but a set of personas related to the personal attributes that users want to test in their privacy hypotheses. This enables users to explore and audit the causal relationship between privacy attributes and the recommended content.

\subsection{Counterfactual Reasoning for Enhancing Data Interpretation}

Counterfactual reasoning is a psychological term that means modifying a factual prior event and then evaluating the consequences of that change~\cite{menzies2001counterfactual}. It has definitions very similar to those of contrastive learning~\cite{chen2020simple} in AI and variation theory~\cite{ling2012variation} in education. Contrastive learning differentiates between similar and dissimilar data points to improve \textit{model} understanding of categories~\cite{chen2020simple}. Variation theory focuses on distinguishing essential from non-essential features of concepts to enhance \textit{human} learning~\cite{marton2014necessary}. Although they have different targets, both aim to deepen the target's understanding through comparison and contrast.

Previous work used counterfactual reasoning to improve the explainability of models through heterogeneous information networks~\cite{ghazimatin2020prince}, influence functions~\cite{tran2021counterfactual}, and perturbation models~\cite{xu2021learning}. For example, Ghazimatin et al.~\cite{ghazimatin2020prince} generated explanations from the provider's perspective by identifying a minimal set of the user's past interaction (e.g., reviews, purchases, and ratings) that, if removed, would alter the recommendation results. Xu et al.~\cite{xu2021learning} proposed a perturbation model to generate alternative user histories and recommendation items, then applied a causal rule mining algorithm to identify personalized causal relationships for the recommendation system.

Our approach leverages counterfactual reasoning to highlight the connection between private data and online content recommendation. By creating personas with varied privacy attributes, our system demonstrates the impact of privacy data usage on recommendation outputs. Unlike prior methods that relied on public datasets for validation, our system uses the generated personas' data to replace real user data, enabling tests on actual websites. This enables users to evaluate the effects of their data usage by themselves, promoting a greater sense of control and privacy awareness.

\subsection{Improving Algorithm Accountability in Online Behavioral Advertising}
Online behavioral advertising (OBA) is ``the practice of monitoring people’s online behavior and using the collected information to show people individually targeted advertisements~\cite{boerman2017online}''. Online behavior can include web browsing data, search histories, media consumption data (e.g., videos
watched), app usage, purchases, click-through responses to ads, and communication content via email or social media. Although users and regulations require ad companies to be transparent about privacy data processing practices~\cite{gomez2009knowprivacy, regulation2016regulation}, Many users are still concerned about the unwarranted use of their data in the OBA~\cite{ur2012smart, varnali2021online, boerman2017online}. 

Informed consent and privacy policy statements serve as the two widely used tools to improve OBA transparency. Although such tools explain the collection and usage of private data, people often cannot find them due to dark patterns~\cite {lu2024awareness} or merely ignore them due to their complex content~\cite{rudolph2018users}. To mitigate this issue, Van et al. added descriptive information to the privacy icon in OBA, indicating that the ad is shown based on the personal online behavior of users~\cite{van2013online}. Compared with these static approaches, we focus on creating an interactive system to allow users to examine different influences of privacy attributes on online ads. 
Enhancing user engagement in auditing their privacy settings is crucial because it empowers individuals to actively understand and control how their data is used. Experiential learning~\cite{kolb2014experiential} shows that interaction fosters deeper comprehension of complex systems like online behavioral advertising. By allowing users to explore and verify privacy hypotheses, they gain a more intuitive understanding of privacy risks and outcomes, increasing their sense of control and trust.
Therefore, our approach enables users to learn from exploration, where they can interactively verify their privacy hypothesis and observe the causal relationship between privacy attributes and system outcomes. 


\section{The Privacy Auditing Sandbox Approach}
To democratize users' privacy auditing for online content recommendation, we introduce a novel privacy auditing sandbox approach. This method allows end users to conduct real-time audits of the relationship between personal user attributes and online content recommendations---a capability previously unavailable. 
Prior work has suggested an approach to generate realistic privacy personas using few-shot learning, contextualization, and chain-of-thoughts techniques~\cite{chen2023empathy}. Build on this approach, our method enhances user capabilities by enabling them to: (1) generate a set of personas and corresponding personal data based on controled variaions in specified attributes; (2)import this synthetic personal data into browsers for the automated collection of online content; and (3) observe the relationship between recommended contents and personal attributes to verify users' hypotheses. 

The key goals of our privacy auditing sandbox approach are two-fold:
\begin{enumerate}
    \item \textbf{Support users to explore the relationship between individual user attributes and recommended content.} When users observe a correlation between personas' personal attributes and recommended content (e.g., ``low-income personas are more likely to see coupon or discount ads, while high-income users are more likely to see luxury ads''), they can confirm the influence of users' personal attribute on the online content recommendations.
    \item \textbf{Enhance users' engagement and initiative in privacy audits.} When users hypothesize that certain characteristic (e.g., age) influences online content recommendations, they can take the initiative to create a set of personas with variations in the specific personal attribute (e.g., young, middle-aged, and old personas) and observe whether the recommended content changes accordingly. This approach is inspired by experiential learning~\cite{gentry1990experiential}, allowing users to quickly experience the impact of different personal attributes on recommended content. We believe that this privacy auditing sandbox approach can democratize privacy audits for end-users and enhance their engagement in privacy audits.
\end{enumerate}




Our approach consists of two primary steps: persona variation and hypothesis consolidation (Fig.~\ref{fig:overall_approach}). In the persona variation phase, the system generates a set of synthetic personas based on the privacy attributes users wish to test, ensuring these personas reflect realistic variations. Subsequently, during the hypothesis consolidation phase, users deploy these personas on specific websites to collect online recommendations and analyze their association with the privacy attributes. 

For the scope of this paper, we chose online advertisements as an example domain for online content recommendation due to their pervasive impact on daily life and their relevance to privacy concerns. 
As outlined in previous studies on online behavior modeling~\cite{yao2017folk}, there is a disconnect between the black-box nature of targeted online ads and users' mental models of how these ads operate. While many users may suspect that their personal data and online behavior shape the advertisements they see, they often lack the tools to verify these suspicions. Future research could extend to other algorithmic systems, such as social media feeds, personalized news recommendations, and more. The approach we introduce should be applicable to a broad range of domains. 

In the subsequent sections, we will elaborate on the subject of our privacy audit, describe two pivotal features of our approach, and introduce a prototype system that integrates these features---the Privacy Auditing Sandbox.

\begin{figure*}
    \centering
    \includegraphics[width=\linewidth]{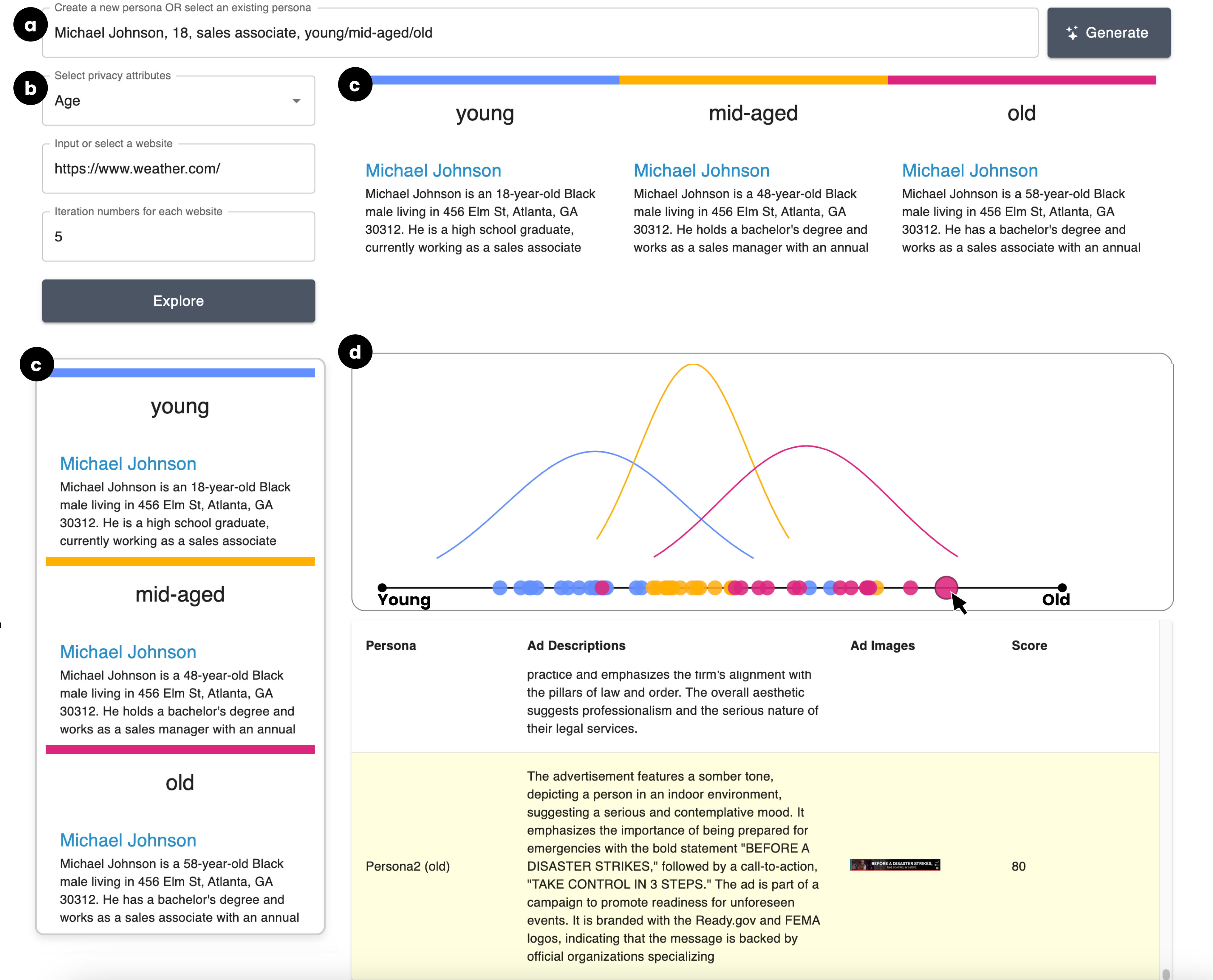}
    \caption{\revise{System Overview: (a) An input field for generating the base persona profile. (b) A control panel to select personal attributes for persona variants, choose a website to audit, and set visit frequencies. (c) A display showing selected personal attribute values, along with the names and descriptions of color-coded persona variants. As users scroll, persona variants shift to the left for continued reference. (d) Visualization of ad distribution based on rated scores, with a scatter plot where each point represents an ad. The ads in the image were obtained by visiting https://www.thepioneerwoman.com/ twice using a persona set consisting of 3 variants. The X-axis shows the Ad-Attribute Alignment score, and the Y-axis shows score probability density. Points are color-coded by persona variants. Users can hover over a point to view the corresponding persona, ad image, description, and rating in the ad list below.}}
    \label{fig:system overview}
    \Description{This figure demonstrates the system overview of Privacy Auditing Sandbox, including four parts: (a) An input field for generating the base persona profile. (b) A control panel to select personal attributes for persona variants, choose a website to audit, and set visit frequencies. (c) A display showing selected personal attribute values, along with the names and descriptions of color-coded persona variants. As users scroll, persona variants shift to the left for continued reference. (d) Visualization of ad distribution based on rated scores, with a scatter plot where each point represents an ad. The ads in the image were obtained by visiting https://www.thepioneerwoman.com/ twice using a persona set consisting of 3 variants. The X-axis shows the Ad-Attribute Alignment score, and the Y-axis shows score probability density. Points are color-coded by persona variants. Users can hover over a point to view the corresponding persona, ad image, description, and rating in the ad list below.}
\end{figure*}





\subsection{Subject of the Privacy Audit}

We clarify that the subject of the privacy audit in our method is the relationship between user characteristics and online content recommendation. 
The term 'user characteristics' refers to a specific trait reflected by the user's online profile and personal data (such as the user's age group as inferred from their account profile and browsing history). 
Note that the subject of our privacy audit is not any particular piece of personal data (such as a specific age-related data item in the account profile or browsing history). In other words, we are not auditing whether the system collects a specific piece of data, nor are we aiming to show users how particular personal data contribute to online content recommendations. This is because current content recommendation systems are complex and opaque black-box systems, making it impossible for us to trace how specific personal data is collected and used by such systems. Therefore, our method focuses on user characteristics as reflected by multiple pieces of personal data, enabling privacy audits by allowing users to observe how system outputs, in the form of recommendations, change in response to changes in user characteristics.

\begin{figure*}
    \centering
    \includegraphics[width=\linewidth]{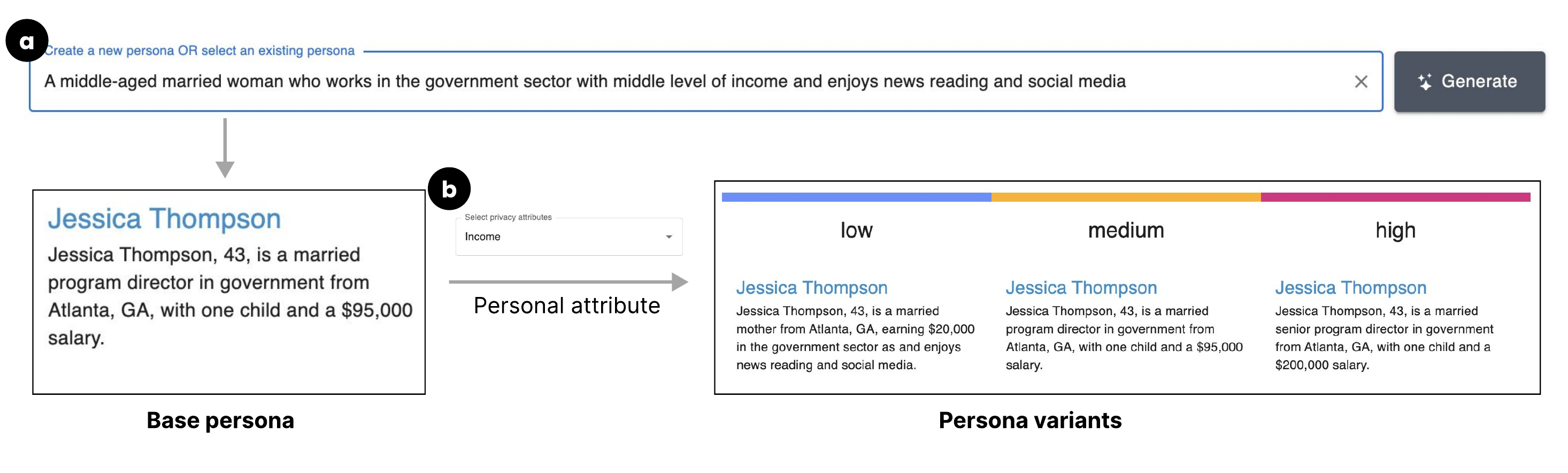}
    \caption{Persona variation. (a) Providing guidance for base Persona’s profile generation. (b) Selecting personal attribute to generate persona variants.}
    \Description{This figure demonstrates the process of persona variation. Part (a) shows how a base persona is created by providing guidance in the form of a descriptive profile, which in this example represents Jessica Thompson, a middle-aged, married woman working in government. Part (b) illustrates how personal attributes, such as income, are selected to generate persona variants. The resulting variants include low-income, medium-income, and high-income versions of Jessica Thompson, each with differing financial details but consistent core characteristics, highlighting how personal attributes are manipulated to create variations for privacy testing.}
    \label{fig: persona variation}
\end{figure*}

\subsection{Persona Variation}
\label{Sec: persona variation}

As defined by prior research~\cite{chen2023empathy}, the privacy persona represents a fictional user with a distinctive biography, demographic information, and a large set of synthesized personal data. Therefore, persona generation is to create a single artificial persona containing realistic synthetic personal data. Compared with it, the goal of persona variation is to create a set of synthetic personas with corresponding personal data that represent variations in a chosen personal attribute. \revise{Our work uses an established persona creation methodology from prior research~\cite{chen2023empathy}.} When users engage with the sandbox, it utilizes synthetic persona data instead of actual user data, enabling testing of how online content recommendations adapt to changes in persona characteristics. By comparing the recommended content with each variant, users can confirm their hypotheses about the link between user characteristics and content recommendations.

Our method begins with a base persona, derived from user input and a selected personal attribute. Using the specified prompt, we generate descriptions for persona variants. Drawing on prior research in persona generation techniques~\cite{chen2023empathy}, we construct comprehensive persona profiles that include demographic details, additional personal information, and longitudinal data. As illustrated in Fig.~\ref{fig: persona variation}, these variants maintain the core personal profile of the base persona but vary based on selected personal attributes. To ensure realistic profiles, not only the chosen attribute but also other related attributes, may be adjusted coherently. For instance, if income level is the selected attribute, the persona's job title might also be modified within the same career path to match the new income bracket.

\promptbox{
\textbf{Prompt for generating persona variants}

Given a fictional user's description: \{\textcolor{brown}{base persona}\} Modify the description to make it more like a \{\textcolor{brown}{privacy attribute}\} version of the person. Do not change age, gender, location, income, and educational level except for the \{\textcolor{brown}{privacy attribute}\}. Only make necessary modifications. In the result, the information about the \{\textcolor{brown}{privacy attribute}\} must appear. Return the profile in only one paragraph.
} \label{prompt for persona variant}

\begin{figure*}
    \centering
    \includegraphics[width=\linewidth]{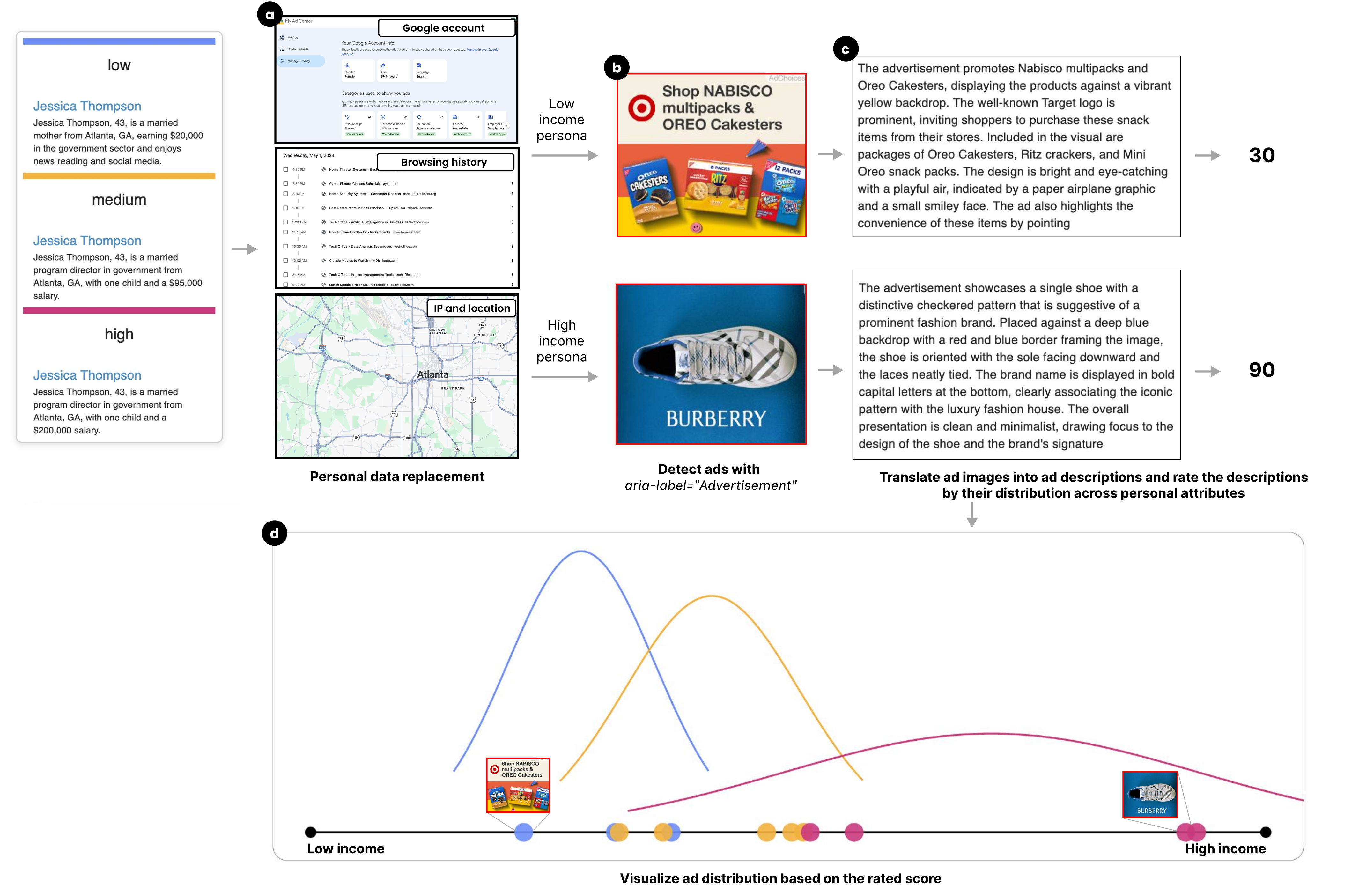}
    \caption{Hypothesis consolidation (a) Replacing the user's real data with each persona variant's data, including Google account, location, IP address, user agent, and browsing history. (b) Detecting online ads displayed on the selected website. \revise{(c) Translating ad images into textual descriptions and rating them on a Ad-Attribute Alignment Score to evaluate their alignment with personal attributes. (d) Visualizing the ad distribution based on the rated scores. Each point on the scatter plot represents an ad, with the X-axis indicating the Ad-Attribute Alignment Score and the Y-axis representing the probability density of the scores based on a normal distribution curve. Points are color-coded by one of the three persona variants.}}
    \Description{This figure demonstrates the hypothesis consolidation process, including four parts: (a) Replacing the user's real data with each persona variant's data, including Google account, location, IP address, user agent, and browsing history. (b) Detecting online ads displayed on the selected website. (c) Translating ad images into textual descriptions and rating them on a Ad-Attribute Alignment Score to evaluate their alignment with personal attributes. (d) Visualizing the ad distribution based on the rated scores. Each point on the scatter plot represents an ad, with the X-axis indicating the Ad-Attribute Alignment Score and the Y-axis representing the probability density of the scores based on a normal distribution curve. Points are color-coded by one of the three persona variants.}
    \label{fig: hypothesis consolidation}
\end{figure*}

\subsection{Hypothesis Consolidation}
\label{Sec: hypothesis consolidation}
At this stage, users substitute their real data with the generated persona data to collect content recommendations as the persona and assess how variations in persona characteristics impact these recommendations. For each persona variant, the privacy auditing sandbox substitutes the user's real data with synthetic data matching the persona’s traits. Thus, when an online service requests this personal information, it will receive the synthetic data corresponding to each persona variant. The system provides an easy-to-use interface that allows users to browse and compare content recommendations received under different personas, thereby facilitating the validation of their hypotheses for auditing.

\textbf{Persona data replacement.} The sandbox comprehensively replaces the user data with the data of each persona variant on Google accounts, location, IP address, user agent, and browsing history, as shown in Fig.~\ref{fig: hypothesis consolidation}a. For Google accounts, it replaces data in the Google Ad Center by creating a dedicated Google account and automating profile updates using the Puppeteer libraries~\footnote{\url{https://pptr.dev/}}. This involves identifying and replacing specific attributes (age, gender, etc.) with persona data in three steps: accessing the profile page, extracting persona attributes based on the alias label in the source code, and replacing the values. For geographical location, it uses OpenStreetMap’s API to get the persona’s latitude and longitude, then override the location in Chrome via Puppeteer. For IP address, it uses NordVPN's API to connect to the server closest to the synthesized location of the persona, ensuring that online services see the server’s IP instead of the user’s. For user agent information, which is a characteristic string that lets servers and network peers identify the application, operating system, vendor, and/or version of the requesting user agent, it uses Puppeteer’s ``setUserAgent'' to replace current details with those that match the persona. For browsing history, it overwrites Chrome's local SQLite database with the persona’s browsing data before launching the browser. \revise{Specifically, both the URL table (which logs visited links) and the visit table (which records timestamps) are replaced with the browsing history of the generated persona. All data from the persona variant, including Google account details, location, IP address, user agent, and browsing history, are generated by LLMs. This ensures consistency and alignment with the persona's synthesized characteristics, allowing a comprehensive and controlled evaluation of the system's responses to the various attributes of the persona.}

\textbf{Online Ad identification.} After loading the persona variant's information into the browser, the sandbox automatically visits the user-specified websites for the number of iterations selected to scrape ads. As shown in Fig.~\ref{fig: hypothesis consolidation}b, if online ads are detected during the visit, the system takes screenshots and stores them in base64 format. Online ads share a common characteristic---their aria-label\footnote{\url{https://developer.mozilla.org/en-US/docs/Web/Accessibility/ARIA/Attributes/aria-label}} of the corresponding HTML element should be ``Advertisement'' based on Web Accessibility Standards. Consequently, when a DOM element with this label is detected on the page, the system identifies its location and takes a screenshot.


\textbf{Ad translation and rating.} To account for both the visual elements and semantic information when evaluating ads against personal attributes, the system first uses GPT-4V to convert each captured ad image into a textual description. \revise{As shown in Fig.~\ref{fig: hypothesis consolidation}c, GPT-4 analyzes these textual descriptions to assess how well each ad aligns with specific personal attributes using an Ad-Attribute Alignment Score.
We define the Ad-Attribute Alignment Score as a metric ranging from 0 to 100 on how well a given ad description aligns with a specific persona attribute (e.g., age, gender, income). A lower score indicates the advertisement is more aligned with one end of the attribute spectrum (e.g., young for age, lower income level for income), while a higher score suggests stronger alignment with the constrasting end (e.g., older people for age, higher income for income). In its implementation, we request the LLM to consider the percentile distribution of the ad (in all ads) ordered by how much it specifically targets either extreme end of the attribute spectrum. For example, on the age attribute, an ad that specifically targets a young audience \textit{the most} would receive close to 0, and an ad that specifically targets a old audience \textit{the most} would receive close to 100.}

\revise{The assessment considers contextual clues such as the advertised product, the brand, the tone of the message, and visual indicators. For instance, an ad for luxury watches might score high on the income scale, indicating alignment with high-income personas, while an ad for budget groceries would score low, aligning with low-income personas. The accuracy of this mechanism is evaluated in Section~\ref{sec: eval of ad rating}.}


\revise{\textbf{Ad distribution visualization.} To help users explore the relationship between recommended ads and persona variants, thereby validating their privacy-related hypotheses, the system visualizes the distribution of ads based on their relevance to the persona attribute used in creating the variants. As shown in Fig.~\ref{fig: hypothesis consolidation}d, the system plots ads on a one-dimensional scatter plot along a coordinate axis based on their Ad-Attribute Alignment Scores. Each point represents an ad, with its position on the x-axis encoding its score. 
The color of each point corresponds to the persona variant active when the ad was retrieved, making it easy to differentiate the ads’ relevance across personas. To make it easier for users to understand the result, the system overlays the scatter plot with a normal distribution curve, allowing users to intuitively observe patterns in ad distributions.} If users want to view detailed information about a specific ad, they can hover over the ad point and the corresponding persona variant, ad image, ad description, and rating will be displayed in the ad list below for users' reference.

\subsection{Example Use Case} We illustrate the functionality of the Privacy Auditing Sandbox through a specific example use case.
In this scenario, a user, Alice, aims to determine if income data affects online advertisements on an e-commerce website~\footnote{https://www.thepioneerwoman.com/}. Alice employs the Sandbox to create various personas, each with distinct annual income levels. She then uses the Sandbox to repeatedly automatically visit the website with personal data associated with each persona and collect data on the advertisements displayed to each. This method allows her to investigate her hypothesis that the website may be targeting ads based on user income, showcasing the core capabilities of the Privacy Auditing Sandbox.

\begin{enumerate}
    \item \textit{Providing Guidance for Base Persona Generation}: 
    Alice enters her guidance to generate a persona that is similar to her profile. She enters ``A middle-aged married woman who works in the government sector with a middle income and enjoys news reading and social media'' (Fig.~\ref{fig: persona variation}a). This guidance serves as the foundation for creating the base persona.  Users can provide as little or as much information as they feel comfortable with, ensuring flexibility while protecting their own private information. Typically, using a base persona that is similar to the real profile of the user fosters empathy building and allows the user to better interpret the results in familiar contexts, as reported in~\cite{chen2023empathy}. The information provided to generate the base persona is not restricted to age or gender but could include marital status, interests, or any other relevant information.  Upon processing Alice's guidance, the Privacy Auditing Sandbox generates a base persona: Jessica Thompson, a 43-year-old married woman living in Atlanta. She works as a program director associate and earns an annual income of \$95,000. She enjoys news reading and social media. Alice can then review and, if necessary, update the guidance to regenerate the persona.
   
    \item \textit{Selecting Personal Attribute for Persona Variation}: 
    Once Alice finalizes the base profile, she selects ``Income'' as the attribute for persona variation.  The Privacy Auditing Sandbox then creates three variants of Jessica Thompson based on the original profile—each representing low, medium, and high-income levels as illustrated in Fig.~\ref{fig: hypothesis consolidation}a. To enhance the realism and consistency of these variants, changes in the selected attribute (income) lead to corresponding adjustments in other personal details, such as job titles.  For each version of the persona variant, the Privacy Auditing Sandbox also generates detailed personal data including the persona’s device and browser in use, home location, weekly schedule with location records, and browsing history. These data are tailored to align consistently with each variant’s specific income profile.
    
    \item \textit{Collecting Online Ads with the Generated  Persona Variants}: 
 Alice inputs a URL for the auditing target and specifies the number with which the website will be sampled for each persona variant. The Privacy Auditing Sandbox then activates a Chrome extension that substitutes Alice's real privacy data with the synthesized data of different variants of Jessica.  Alice’s privacy data, including her profile in the Google Ad Center, browsing history, real-time location, and IP address, are temporarily replaced. The Privacy Auditing Sandbox will automatically visit the target URL repeatedly according to the number of samples, identify online ads, and take screenshots.

    \item \textit{Reviewing the results to validate the hypothesis}: After collecting all the advertisements, the Privacy Auditing Sandbox evaluates each ad by assigning it a score based on its percentile ranking related to the selected personal attribute, ``income'' (Fig.~\ref{fig: hypothesis consolidation}c). Finally, the ads will be visualized on a distribution chart and an ad table based on the scores (Fig.~\ref{fig: hypothesis consolidation}d). Alice reviews these visualizations to analyze the correlation between the income levels of the persona variants and the pricing of products or services featured in the ads. This analysis helps her validate the hypothesis: higher-income personas are more frequently targeted with advertisements for luxury goods on this website.
\end{enumerate}

\subsection{Implementation}
We developed the Privacy Auditing Sandbox with
a frontend using the framework Next.js and a backend powered by Flask. They communicate through HTTP requests for API access.

For data storage, we utilize JSON files to manage the synthesized personal privacy data and the corresponding advertisements collected through the personas. To create synthetic personas, we enhanced our methodology by integrating a Python script that interfaces with the GPT-4 public API using the ``langchain'' Python library, building on the foundational work cited in \cite{chen2023empathy}.

\section{Technical evaluation}
\revise{We conduct a technical evaluation of the Privacy Auditing Sandbox to assess the following technical research questions (TRQs):}

\revise{\textbf{TRQ1:} What is the quality of LLM-generated persona variants (described in Section~\ref{Sec: persona variation})  in terms of contextual relevance, accuracy, hallucination, and stereotyping?}

\revise{\textbf{TRQ2:} How accurate is the ad identification mechanism (described in Section~\ref{Sec: hypothesis consolidation}) ?}

\revise{\textbf{TRQ3:} How stable is the Ad-Attribute Alignment Score (described in Section~\ref{Sec: hypothesis consolidation}) across different persona attributes and ad contexts? }

\revise{\textbf{TRQ4:} What is the impact of persona substitution on the personalized ad content?}
\begin{itemize}
    \item \revise{\textbf{TRQ4a:} Do similar personas, based on chosen attributes, on the same website receive similar ads?}
    \item \revise{\textbf{TRQ4b:} Do different personas, based on chosen attributes, on the same website receive different ads?}
\end{itemize}


\begin{figure}
    \centering
    \includegraphics[width=0.7\linewidth]{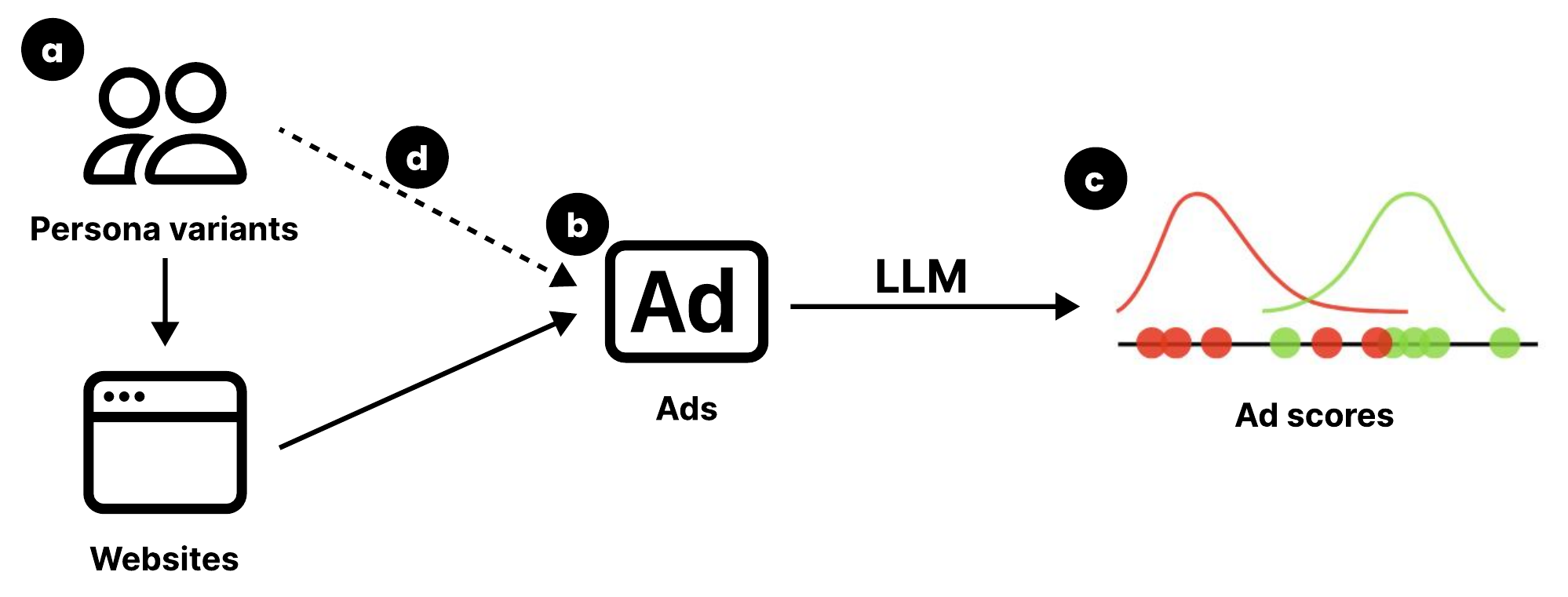}
    \caption{\revise{An overview of our technical evaluation. (a) Conducted an expert review to evaluate the quality of the generated personas. (b) Evaluated the accuracy of ad identification. (c) Assessed the stability of LLM-generated ad scores based on privacy attributes. (d) Investigated the impact of persona substitution on the ad content.}}
    \Description{This figure shows an overview of the technical evaluation including four parts: (a) Conducted an expert review to evaluate the quality of the generated personas. (b) Evaluated the accuracy of ad identification. (c) Assessed the stability of LLM-generated ad scores based on privacy attributes. (d) Investigated the impact of persona substitution on the ad content.}
    \label{fig: technical evaluation overview}
\end{figure}

\subsection{\revise{Overview}}
\revise{As shown in Fig.~\ref{fig: technical evaluation overview}, the Privacy Auditing Sandbox consists of three primary stages: (1) generating persona variants based on specific privacy attributes; (2) substituting user data with persona-specific data on target websites; (3) identifying and collecting advertisements on the target websites; and (4) employing an LLM to evaluate the advertisements based on the privacy attribute defined in the first stage. }

\revise{To assess the reliability of our approach, we conducted technical evaluations at each stage. In the generation of personas, we performed an expert review to detect potential hallucinations and stereotyping in the generated personas. In the identification of online ads, we evaluated the accuracy of our ad identification mechanism to ensure it correctly detected and classified advertisements across various websites. In the ad rating stage, we assessed the consistency of LLM-generated ad scores to ensure stable and reliable assessments. Lastly, in persona substitution, we analyzed the impact of persona substitution on the types of advertisements generated, shedding light on the relationship between persona attributes and ad content.}

\subsection{\revise{Evaluation of Persona Generation Quality}}

\revise{LLM-generated personas are often criticized for issues such as hallucinations and stereotyping~\cite{choi2024proxona, schuller2024generating}, which can undermine their effectiveness and reliability. While most existing automated metrics assess the performance of LLM models~\cite{min2023factscore, chern2023factool, wei2024long, cheng2023marked}, they inadequately address the quality of generated personas. This gap highlights the need for human evaluation to ensure accuracy and relevance. To address this challenge, we adopted an expert review approach to evaluate persona quality. The evaluation used established scales~\cite{salminen2020persona, salminen2024deus} to assess the quality of personas from the perspectives of consistency, credibility, and stereotypicality.}

\begin{figure*}[h]
    \centering
    \includegraphics[width=\linewidth]{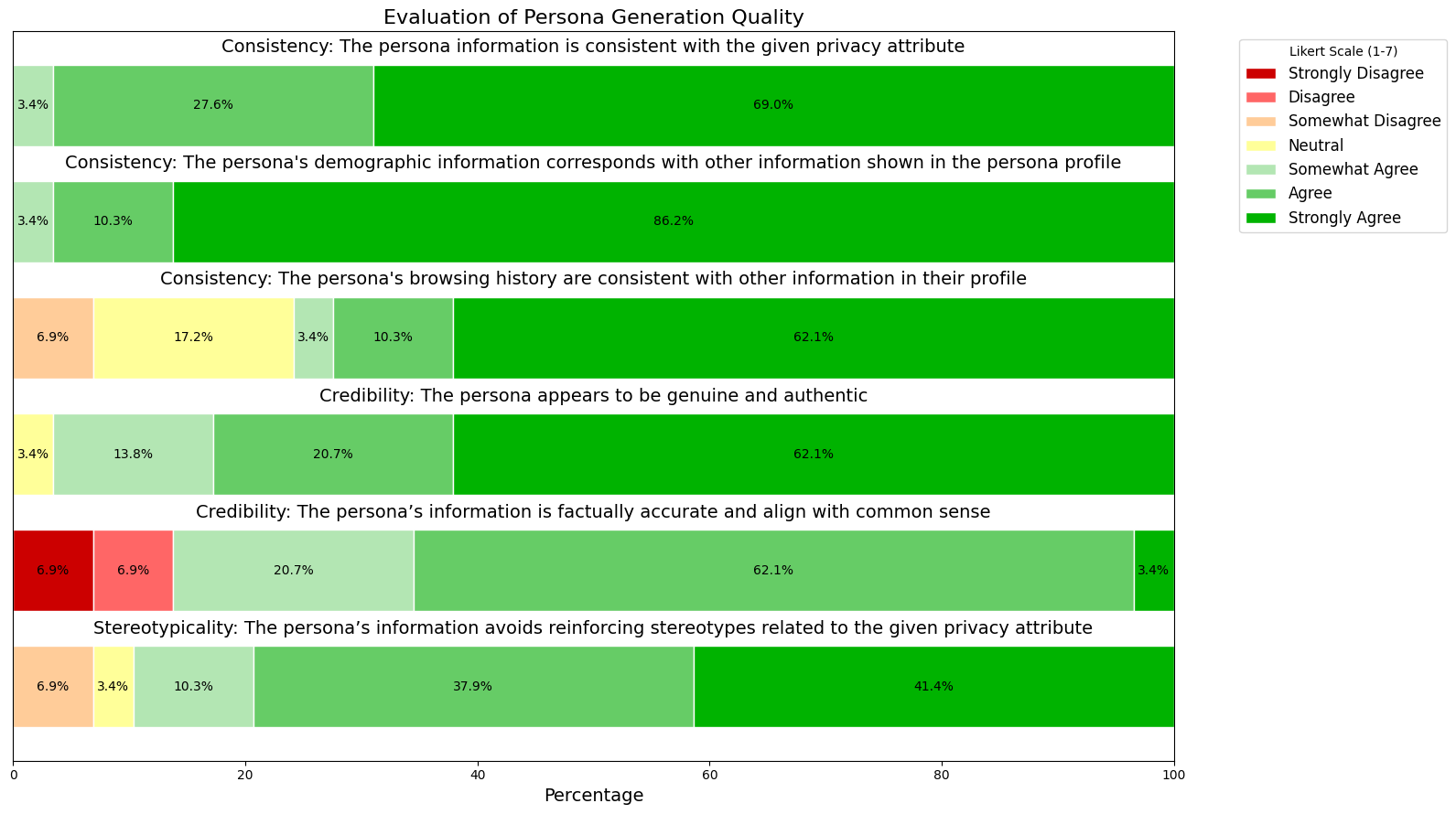}
    \caption{\revise{Result of the expert evaluation on persona generation quality.}}
    \Description{This figure shows a stacked horizontal bar chart titled "Evaluation of Persona Generation Quality," assessing various aspects of persona generation using a Likert scale from "Strongly Disagree" (red) to "Strongly Agree" (dark green). The highest agreement is observed for "Consistency: The persona's demographic information corresponds with other information" (86.2\% Strongly Agree), while "Credibility: The persona's information is factually accurate" shows mixed responses, with 62.1\% Strongly Agree but 13.8\% disagreement. For "Stereotypicality," most respondents (79.3\%) agree or strongly agree that the persona avoids stereotypes. The chart highlights strong overall agreement on consistency and credibility, with occasional disagreement on factual accuracy and stereotypes.}
    \label{fig: persona_quality}
\end{figure*}

To address \textbf{TRQ1}, we engaged two experts with relevant experience in persona creation to evaluate the quality of generated personas. None of the experts is an author on this paper. Both experts have at least 5 years of experience in UX design and research, are familiar with the use of personas in the UX practice, and have at least a master degree in HCI or relevant fields. Expert reviewers offer nuanced insights into stereotyping and contextual relevance, aspects often overlooked by automated methods. To reduce subjective bias, we implemented a structured evaluation framework combining Likert-scale ratings and open-ended questions to capture qualitative feedback. Additionally, we computed Cohen’s Kappa to measure inter-rater agreement, ensuring reliability and consistency in the evaluation outcomes.

\revise{As shown in Fig.~\ref{fig: persona_quality}, the expert review (Cohen’s Kappa=$0.88$, indicating strong inter-rater reliability) demonstrated that the generated personas exhibited strong consistency with the given privacy attribute. The demographic information of the generated personas was also highly consistent with other information shown in the persona profile. However, there were minor inconsistencies in the browsing history relative to other profile information. This could be due to the dynamic nature of browsing data, which might not always align closely with static persona attributes. In terms of credibility and stereotypicality, most of the generated personas achieved a score of 6 or above on a 7-point Likert scale, meeting baseline criteria for capturing the intended privacy attribute without introducing significant inaccuracies or biases. However, hallucination errors, such as implausible income levels and physical addresses, were identified in a small subset of the personas. These hallucinations may have occurred due to the LLM's lack of understanding of specific real-world geographical or income data, leading to the generation of a small portion of data that did not align with reality.}

\subsection{\revise{Evaluation of Ad Identification}}
\revise{To answer the \textbf{TRQ2} for the accuracy of ad identification, we tested our ad identification module on 205 websites with 805 advertisements. Since there is no publicly available dataset of ad-containing websites, we curated our dataset using a publicly available Kaggle dataset\footnote{https://www.kaggle.com/datasets/hetulmehta/website-classification} of 1,408 websites. Among the 1,408 websites, we manually visited each website’s homepage and identified 205 websites with ads. For those 205 websites, we recorded the positions of 805 displayed advertisements. Our dataset emphasizes the location of advertisements rather than their content. For ads occupying the same position but displaying rotating or refreshing content, we treated them as a single advertisement due to their consistent placement.}

\revise{Using this curated dataset, we evaluated our ad identification module and summarized the results in Table~\ref{table:confusion_matrix}. The method achieved achieved a high accuracy rate of 96.52\%, , with a precision of 100\% (all identified ads are indeed ads) and a recall of 96.52\%, demonstrating its ability to detect the majority of ads across diverse website designs. Errors were primarily associated with dynamically loaded advertisements or non-standard labeling practices on certain websites, which limited detection in a small number of cases. Additionally, some missed detections were due to advertisements embedded in floating window, which presented unique challenges for identification. We plan to address these edge cases in future work.}

\begin{table}[ht]
\small
\centering
\begin{tabular}{p{0.7cm} p{0.7cm} p{0.7cm} p{0.7cm} p{1cm} p{1cm} p{1cm}}
\toprule
\revise{\textbf{TP}} & \revise{\textbf{FN}} & \revise{\textbf{TN}} & \revise{\textbf{FP}} & \revise{\textbf{Acc.}} & \revise{\textbf{Prec.}} & \revise{\textbf{Rec.}} \\
\midrule
\revise{$777$} & \revise{$28$} & \revise{$0$} & \revise{$0$} & \revise{$96.52\%$} & \revise{$100\%$} & \revise{$96.52\%$} \\
\bottomrule
\end{tabular}
\caption{\revise{Performance of the Ad Identification Module. TP = True Positive, FN = False Negative, TN = True Negative, FP = False Positive, Acc. = Accuracy, Prec. = Precision, Rec. = Recall.}}
\label{table:confusion_matrix}
\end{table}

\subsection{\revise{Evaluation of Ad Rating Stability}}
\label{sec: eval of ad rating}
\revise{To answer the \textbf{TRQ3} for the stability of the Ad-Attribute Alignment Score along five dimensions of privacy attribute (age, gender, urbanization of home location, income level, and education level), we selected 20 advertisements per dimension and computed the Ad-Attribute Alignment Score for each ad five times using the LLM. The LLM did not receive information on which persona was associated with each ad. Instead, its evaluations were solely based on its understanding of the underlying privacy attribute.}

\revise{We measured the consistency of the scores using the averaged standard deviation and the averaged coefficient of variation. Lower values for these metrics indicate higher stability and demonstrate the LLM's ability to deliver consistent evaluations across multiple iterations.}

\revise{The experimental results, presented in Table~\ref{table:LLM_rating}, show that that the average standard deviation for the LLM’s scores is 0.87, with an average coefficient of variation of 3.00\%. This indicates that the LLM demonstrated a high stability in its evaluation of advertisements across repeated scoring attempts.
Overall, these results confirm the stability of the LLM's scoring mechanism for evaluating ads based on privacy attributes.}

\begin{table}[ht]
\small
\centering
\begin{tabular}{p{1cm} p{0.6cm} p{0.8cm} p{0.8cm} p{0.8cm} p{1cm} p{1cm}}
\toprule
& \revise{\textbf{Age}} & \revise{\textbf{Gender}} & \revise{\textbf{Location}} & \revise{\textbf{Income}} & \revise{\textbf{Education}} & \revise{\textbf{All}} \\ 
\midrule
\revise{Avg. std} & \revise{0.57}  & \revise{0.75} & \revise{0.63} & \revise{1.19} & \revise{1.22} & \revise{0.87} \\ 
\revise{Avg. cov} & \revise{0.77\%}  & \revise{7.92\%} & \revise{2.75\%} & \revise{2.06\%} & \revise{1.50\%} & \revise{3.00\%} \\ 
\bottomrule
\end{tabular}
\caption{\revise{Results of the Ad Rating Stability. ``std'' = standard deviation, ``cov'' = coefficient of variation.}}
\label{table:LLM_rating}
\end{table}


\subsection{\revise{Evaluation of the Impact of Persona Substitution on Ads}}


\begin{figure*}
    \centering
    \includegraphics[width=\linewidth]{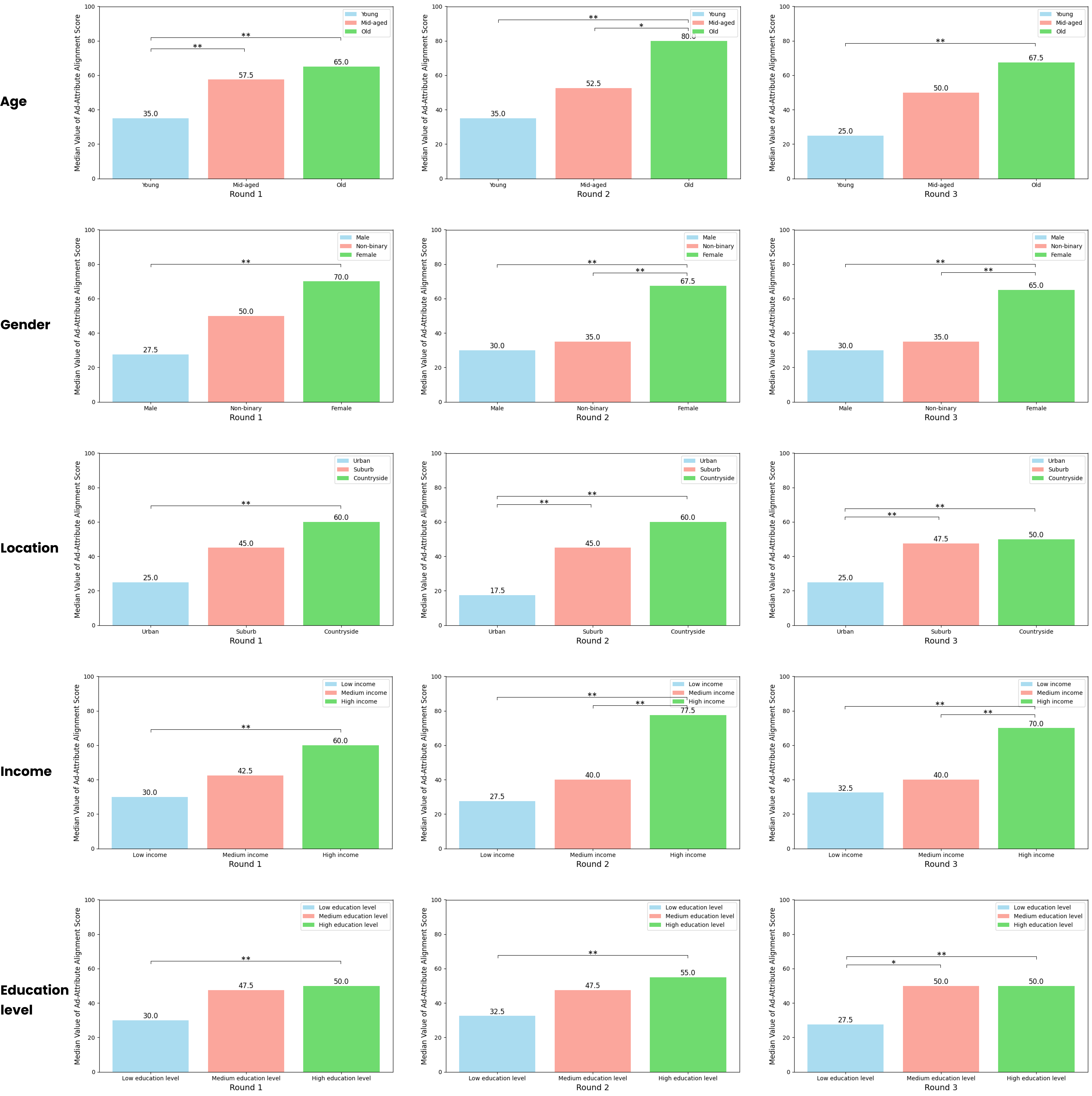}
    \caption{\revise{Comparison of Ad-Attribute Alignment Scores for different personas. Single asterisk (*) denote p<0.1, while double asterisks (**) indicate p<0.05.}}
    \Description{This figure presents a series of bar charts comparing Ad-Attribute Alignment Scores across various personas categorized by Age, Gender, Location, Income, and Education Level over three rounds. Each chart shows three groups within a category, highlighting differences in scores. Across all categories, noticeable trends emerge, such as older individuals, females, rural residents, high-income earners, and those with higher education levels generally achieving higher alignment scores. Statistical significance is marked, with a single asterisk (*) denoting p < 0.1 and double asterisks (**) indicating p < 0.05, emphasizing meaningful differences between groups.}
    \label{fig: different persona}
\end{figure*}

\begin{figure*}
    \centering
    \includegraphics[width=\linewidth]{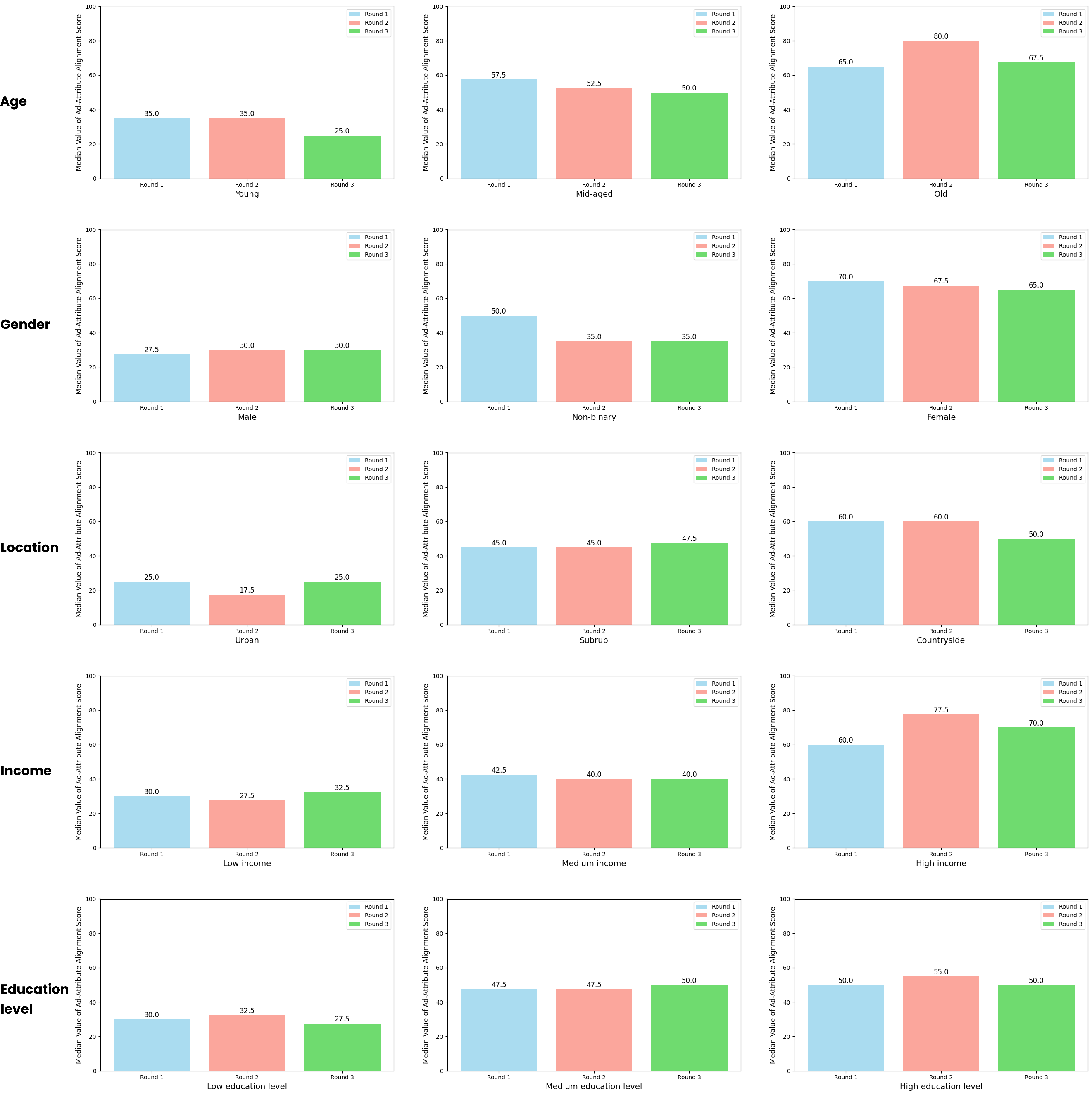}
    \caption{\revise{Comparison of Ad-Attribute Alignment Scores for similar personas. There was no statistical significant difference between any rounds within the same attribute value group.}}
    \Description{This figure presents a series of bar charts comparing median ad-attribute alignment scores across three rounds for personas with varying demographic attributes: age, gender, location, income, and education level. For age, scores show slight decreases for the young and old groups, while mid-aged personas have fluctuating scores. In gender, male and non-binary personas show consistent scores, while female personas experience a slight dip in Round 3. Location shows marginal score increases for urban personas, stability for suburban personas, and slight decreases for those in the countryside. Income personas demonstrate a small increase for low-income groups, steady scores for medium-income, and a noticeable rise for high-income groups in Round 3. Lastly, education level scores fluctuate slightly for low education, increase for medium education, and show a noticeable rise for high education personas in Round 2. Each chart is color-coded by round, with scores ranging from 0 to 100.}
    \label{fig: similar persona}
\end{figure*}

\revise{To address \textbf{TRQ4} regarding the impact of persona substitution on advertisements, we conducted a validation experiment. For each of the five privacy attribute (age, gender, urbanization of home location, income level, and education level), a consistent guidance prompt was used three times to generate three groups of persona variants, 
with each group containing three distinct personas. We identified five representative websites (as detailed in Section~\ref{Sec: websites}). We used each persona to access each one of the five websites, visiting each site three times. During these visits, the user data was substituted with persona-specific data (as described in Section~\ref{Sec: hypothesis consolidation}). For each persona-website pair, we collected and analyzed the advertisements displayed to assess the effect of persona substitution on ad recommendations.}

\revise{Since in the previous subsection we verified the stability of LLM in ad ratings, we proceeded to use the LLM to rate the ads based on their alignment with privacy attributes. Through the ratings, we quantified how much the persona attributes influenced the advertisement content. For each group of three privacy variants (e.g., young, mid-aged, old for age), we treated them as distinctly different personas. For the three personas generated in the same group based on the same privacy attribute values, we treated them as similar personas. For both different personas and similar personas, we used the Kruskal–Wallis H test~\cite{mckight2010kruskal} and post-hoc tests~\cite{dinno2015nonparametric} to evaluate whether there were significant differences in ad-attribute alignment scores among personas from different groups (\textbf{TRQ4b}), and whether the ad-attribute alignment scores for similar personas remained consistent across the three rounds (\textbf{TRQ4a).}}


\revise{Fig.~\ref{fig: different persona} shows the scores for different personas in each round, with all personas generated from contrasting privacy attribute values (e.g., young-old, low income-high income) showing significant differences. This indicates that advertisements shown to distinctly different personas exhibited significant differences (\textbf{TRQ4b}). In contrast, Fig.~\ref{fig: similar persona} shows that personas generated from the same privacy attribute values showed stable rating scores, with no significant differences detected. The analysis revealed that advertisements served to similar personas on the same website were more similar to each other (\textbf{TRQ4a}).}

\section{User Study}

We conducted a user study\footnote{The study protocol was reviewed and approved by the IRB at our institution} with 15 participants to evaluate our approach. The study examines the following research questions.

\textbf{RQ1:} How can Privacy Auditing Sandbox support end users to audit online content recommendations by verifying their privacy hypotheses?


\textbf{RQ2:} How does the engagement of the Privacy Auditing Sandbox influence users' attitudes, awareness, and sense of empowerment with online content recommendations?



\textbf{RQ3:} Do users find the Privacy Auditing Sandbox usable and useful?


\subsection{Participants}
We recruited 15 participants through word-of-mouth and social media, all of whom participated in the study virtually via Zoom. Before the study, participants completed a pre-screening survey that collected essential demographic information such as age, gender, state of residence, race/ethnicity, occupation, and education level. We aimed to recruit a group of participants from diverse demographic backgrounds.  The comprehensive demographic details of the participants are outlined in Table~\ref{tab: Participant Demographics}. The age range of participants was from 22 to 51, including five females and ten males. Each participant was compensated \$25 USD for their participation.

\begin{table*}[ht]
\centering
\begin{tabular}{lllllll}
\toprule
\textbf{ID} & \textbf{Gender} & \textbf{Age} & \textbf{Ethnicity} & \textbf{Educational Level} & \textbf{Occupation}\\
\midrule
P1 & Female & 26 & Asian and Pacific Islander & Master's degree & Financial analyst \\
P2 & Female & 21 & Asian and Pacific Islander & Bachelor's degree & Student \\
P3 & Female & 22 & Asian and Pacific Islander & Bachelor's degree & Student \\
P4 & Female & 51 & White or Caucasian & Master's degree & Property  manager \\
P5 & Male & 22 & Black or African American & High school graduate & Freelancer \\
P6 & Female & 24 & White or Caucasian & Bachelor's degree & Student \\
P7 & Male & 31 & White or Caucasian & Bachelor's degree & Project Manager \\
P8 & Male & 25 & Black or African American & Master's degree & Teacher \\
P9 & Male & 30 & Asian and Pacific Islander & Master's degree & Student \\
P10 & Male & 30 & Asian and Pacific Islander & Doctorate degree & Student \\
P11 & Male & 27 & Black or African American & Bachelor's degree & Technician  \\
P12 & Female & 42 & White or Caucasian & Bachelor's degree & Writer  \\
P13 & Male & 25 & Black or African American & Bachelor's degree & Architect  \\
P14 & Male & 30 & White or Caucasian & Bachelor's degree & Program Coordinator  \\
P15 & Male & 22 & Black or African American & Bachelor's degree & Student  \\

\bottomrule
\end{tabular}
\caption{Participant Demographics}
\label{tab: Participant Demographics}
\end{table*}

\subsection{Study Design}
Each study session lasted around 60 minutes. The session consisted of two phases.

\subsubsection{Study procedure.} After completing the informed consent process and receiving a brief introduction to the study, each participant completed the following two phases of the study.
\begin{itemize}
    \item \textbf{Phase 1: Using existing sets of personas to verify privacy hypotheses.} To understand whether and how users could verify privacy hypotheses through our system, we employed a combination of the ``think-aloud'' method~\cite{jaaskelainen2010think} and semi-structured interviews. The experimenter first introduced the Privacy Auditing Sandbox to the users, ensuring they understood how persona variants were generated and how to load them into the browser to receive ads. 
    Each participant was then asked to review five personal attributes—age, gender, urbanization, income level, and education level—where each attribute corresponded to a different set of pre-existing persona variants and the ads collected using these personas in the Privacy Auditing Sandbox's hypothesis consolidation interfaces. These five attributes were specifically chosen because online recommendation systems frequently rely on this data to tailor their offerings~\cite{mohamed2019recommender}.
    As they navigated the personas, participants were instructed to vocalize the connections they observed between the ads and the personal attributes, as well as the reasons they believed these connections existed. During the ``think-aloud'' process, as participants shared their immediate feedback, researchers could interject with follow-up questions or ask for clarifications. A detailed list of these five sets of personas can be found in the Appendix~\ref{appendix: persona}.
    
    \item \textbf{Phase 2: Exploring privacy hypotheses with generated personas.} The goal of this phase is to evaluate the entire workflow of Privacy Auditing Sandbox, study its impact on participants' understanding of online content recommendations, and explore how this understanding influences their future interactions with websites. Since in the previous phase, participants tested the connections between five personal attributes and online ads, they may develop an interest in a particular personal attribute. This phase allows users to experience the process of creating a persona and collecting ads in real time on actual websites, allowing them to further explore the relationship between personal attributes and online ads. First, the participant enters a natural language description of the persona they want to generate and selects a personal attribute to create a set of corresponding personas. Then, the participant was asked to choose a website and specify how many times each persona should browse that site. The Privacy Auditing System triggers the launch of a new browser window, which automatically replaces their real personal data with synthetic persona data and collects ads from the websites visited. Once all the ads had been collected, the participant reviewed the distribution of ads collected using this set of personas in relation to the personal attributes and explained, in a think-aloud manner, how the ads relate to those personal attributes. 
\end{itemize}
   
    At the end of the study session, the participant completed a System Usability Scale (SUS) test~\cite{bangor2009determining} to assess the system's usability. The experimenter also conducted a brief follow-up semi-structured interview based on the participant's responses and the experimenter's observations during the study session, asking the participant to explain the rationales behind their actions and ratings, elaborate on their thoughts on the usability and usefulness of the system, and describe the impacts of using the system on their privacy awareness, attitudes, and future behaviors. 

    \revise{Note that this study does not include a baseline for comparison, because we cannot identify an existing tool that can support end users to verify privacy hypotheses and explore the relationship between personal attributes and online ads in third-party websites in the similar way as the Privacy Auditing Sandbox. Instead, we focus on evaluating the usability, user interaction, and perceived usefulness of the system.}

\subsubsection{Websites.} 
\label{Sec: websites}
We selected five representative websites (as shown in Table~\ref{tab:website_list}) to test the Privacy Sandbox with personalized advertisements. Our website selection process was based on the methodology previously employed by Zeng et al.~\cite{zeng2022factors}. The selection was based on the following criteria: 1) representing a diverse range of website topics, 2) presenting multiple advertisements on each website, and 3) using Google Ads as the primary ad provider. The decision to use Google Ads was driven by two key factors: (1) Google Ads is the most widely used advertising platform on the Internet, reaching over two million websites and apps and more than 90\% of Internet users worldwide~\footnote{https://support.google.com/google-ads/answer/117120?hl=en}; and (2) Google Ads has extensive access to personal data stored in the Chrome browser (e.g., browsing history, Google accounts), which was central to our study.

\begin{table}[ht]
\small
\centering
\begin{tabular}{llr}
\toprule
Website & Topics &  Site Rank \\
\midrule
www.weather.com & Weather forecasts & 35 \\
www.foxnews.com & National news & 43 \\
www.thepioneerwoman.com & E-commerce & 1,049 \\
boardingarea.com & Traveling & 27,905 \\
www.algebra.com & Education & 54,809 \\
\bottomrule
\end{tabular}
\caption{Websites visited by participants in the study}
\label{tab:website_list}
\end{table}

\subsection{Data Analysis Methods}
We employed standard open coding procedures~\cite{brod2009qualitative} to perform qualitative data analysis on the transcripts of user think-aloud sessions and the concluding interviews of the study. Two researchers in our team independently began the coding process using MAXQDA. One researcher coded 20\% of the sample, generating an initial set of codes. The second researcher then coded the same portion, introducing new codes as needed. Any discrepancies between their coding were resolved through discussion to ensure consistency, resulting in a unified codebook. Then, this codebook was used to perform a thematic analysis, identifying and defining key themes that emerged from the interviews and aligned with the established codes.

The complete codebook is presented in Appendix~\ref{appendix: codebook}. These themes were then refined and developed to inform the study's findings, which are presented in detail in Section~\ref{Sec: result}.
\section{Results and Findings}
\label{Sec: result}

\begin{figure*}
    \centering
    \includegraphics[width=\linewidth]{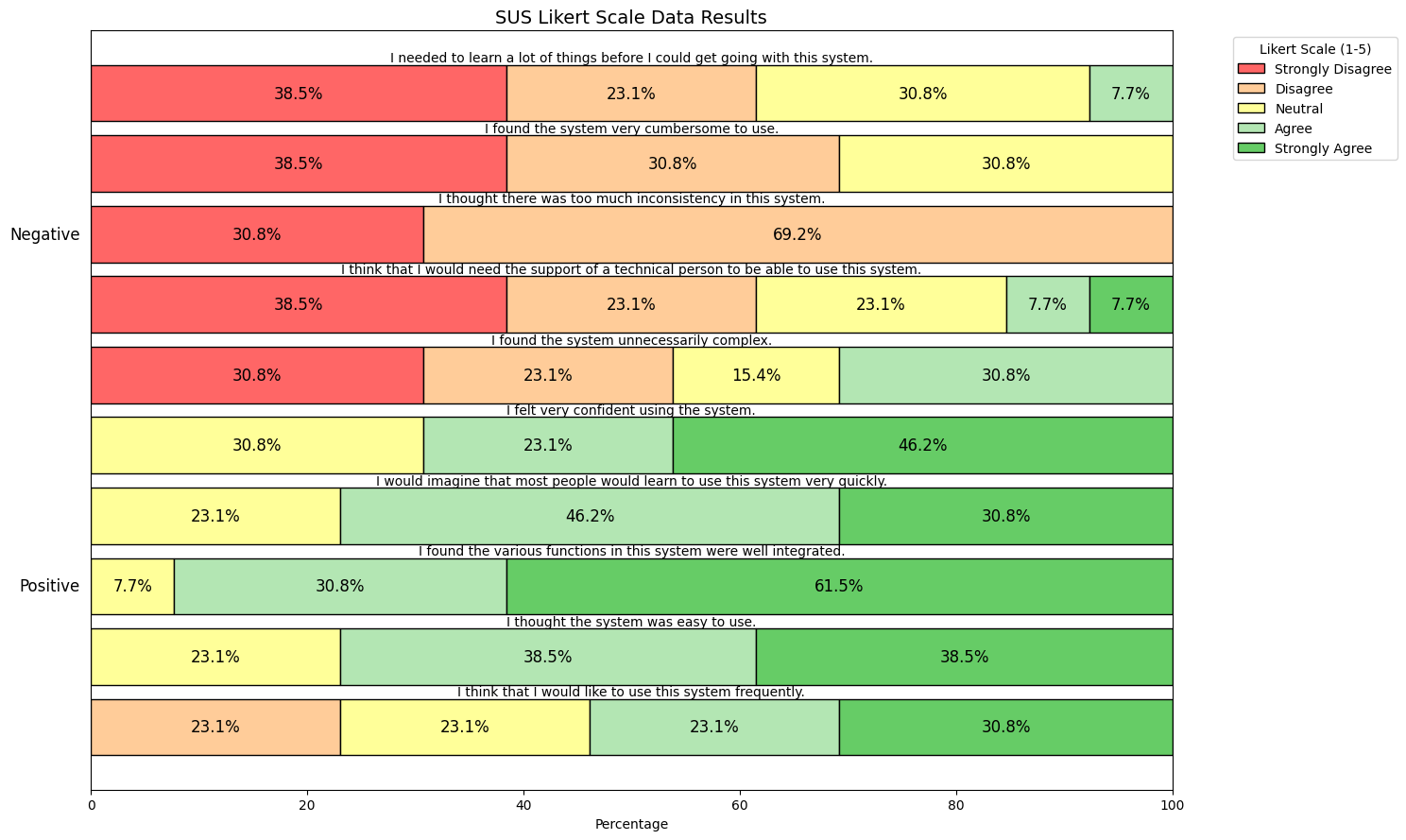}
    \caption{Result of Post-study Questionnaire. The statements are categorized into positive and negative ones based on their wording. \revise{For the negative statements (first five), lower ratings are better. For the positive statements (bottom five), higher ratings are better.}}
    \Description{This figure presents the results of a post-study questionnaire using a System Usability Scale (SUS) Likert scale. The statements are categorized into positive and negative groups, with responses rated on a 5-point scale from "Strongly Disagree" to "Strongly Agree." The negative statements, such as "I needed to learn a lot of things before I could get going with this system," show higher percentages of disagreement (red) and neutrality (yellow), which is preferred for negative statements. Conversely, positive statements like "I felt very confident using the system" show a high percentage of agreement (green), indicating favorable user feedback. The figure highlights that lower ratings are better for negative statements, while higher ratings are better for positive statements, reflecting overall user satisfaction and ease of system use.}
    \label{fig: sus}
\end{figure*}


Our study demonstrates that the Privacy Auditing Sandbox effectively enabled participants to validate their privacy hypotheses by identifying the connections between personal attributes and targeted advertisements. Participants successfully linked personal attributes to ad themes and visual elements, highlighting patterns related to age, gender, and income (\textbf{RQ1}). After using our tool, users expressed discomfort over privacy intrusions(\textbf{RQ2}). The system was generally rated as user-friendly and useful (\textbf{RQ3}).

\subsection{RQ1: Participants validated their privacy hypotheses by linking personal attributes with the theme of the ad and visual elements}
\label{Sec: result rq1}.

\rev{The Sandbox enabled participants to systematically test their assumptions and observe real differences across persona variants, revealing hidden patterns in ad content.} As shown in the result of our post-study SUS questionnaire in Fig.~\ref{fig: sus}, over 75\% of participants agreed that they could quickly learn to use the Privacy Auditing Sandbox. Once they learned it, over 75\% of participants found our system easy to use. \rev{This ease of use was crucial in encouraging users to try multiple hypotheses and interact deeply with the system. All participants discovered that personal attributes clearly influenced ad recommendations. One participant (P5) explained:} , ``\textit{I think it (the Privacy Auditing Sandbox) is pretty easy (to use). You just input the person you're thinking about and then compare the differences (among ads), whether those are based on gender, age, education, or work background.}'' Below we summarized two focus areas where participants linked personal attributes with ads to validate their privacy hypotheses: the theme and visual elements of the ads.

\subsubsection{Ad themes}
In the auditing process, participants successfully validated the impact of persona variants on ads primarily through the correlation between the ad theme and personal attributes, particularly in the areas of age, gender, and income. For age, participants found that young personas were more likely to receive ads for video games, fast food, and workouts, while older personas were shown more ads related to financial aid, healthcare, and skin repair. The users also provided some explanations. For example, P5 said ``\textit{The ad shows a smiling young person playing video game...This one definitely has a relevancy to the age since young people are probably interested in gaming.}''. P1 said, ``\textit{For these (middle-aged and old) people, financial aid is probably what they need, because maybe in middle age, like people who own their own company, they might fall into some bad financial situation.}'' 
In terms of gender, participants believed that ads targeting women were easier to identify. They observed that ads for women often focused on footwear or fitness, while ads targeting men included more tech-related content (e.g., Starlink). Participants did not find any ads specifically targeting non-binary individuals when browsing ads received when using the non-binary persona. Regarding income level, participants noticed that lower-income personas were more likely to see ads for coupons, financial aid, and fast food, while middle- and high-income personas, although occasionally exposed to discount-related ads, were the only ones shown ads for expensive products (e.g., financial products, luxury goods), as P2 said ``\textit{luxury products only appear in the person with very high income.}''

\subsubsection{Visual elements}
The visual elements in advertisements can also help participants identify the connection between ads and persona characteristics, particularly in terms of gender and age. If an ad features a female character or uses colors like pink or purple, participants are more likely to think the target audience is female. For instance, P1 said, ``\textit{The advertisement for females tends to use female models.}'' Similarly, if the ad features an older person, participants assume that the target audience is also older adults. As P7 noted, ``\textit{Even though the ad topic is general, the visual element is an older gentleman, which means that it has some relation with older people.}''

\subsection{RQ2: Participants exhibited conflicting emotions between accepting personalized content recommendations and concerns about privacy invasion}
\label{Sec: result rq2}

\rev{The Sandbox helped participants reflect on their emotional responses to personalized recommendations. This reflection surfaced tensions between appreciation for relevance and discomfort about surveillance.} Some participants were somewhat accepting that websites use personal attributes to tailor advertisements, appreciating that this customization could enhance the relevance of content and facilitate more targeted product or service offerings. P5 said: ``\textit{I feel like I am more influenced to shop at places when I get targeted ads that align with my personal attributes.}'' 
\rev{But many also felt discomfort when seeing how easily systems inferred personal details.} P10 noted: ``\textit{But sometimes it annoys me because I am... let's say I looked at luxury cruises just for fun. And then suddenly they pop up everywhere... The system knows what I'm doing, which bothers me.}'' 
\rev{By making this targeting process visible, our system moved users from vague discomfort to informed concern—making privacy risks tangible and specific.
Some participants expressed ``privacy resignation.''~\cite{draper2017privacy}, knowing what’s happening, but feeling unable to change it. Even in these cases, the tool gave users language and structure to describe what they were experiencing—transforming abstract discomfort into concrete awareness. }



\subsection{RQ3: Participants found the privacy auditing sandbox generally usable and useful}
\label{Sec: result rq3}

\rev{Overall, participants found the tool easy to learn, well-structured, and effective for exploring the impact of personal data on recommendations.}
The System Usability Scale(SUS) questionnaire~\cite{bangor2009determining}, indicates a generally positive user experience, as shown in Fig.~\ref{fig: sus}. Participants rated each item on a 5-point Likert scale. They found the system user-friendly and well-integrated, with most anticipating a shallow learning curve.  
The confidence in using the system is high. 

Participants do not perceive too much inconsistency in the system, nor do they feel cumbersome to use the system. \rev{Participants highlighted that the system helped them ask better questions about data use.} For example, P15 shared, ``\textit{For me, I think it's useful. If, for example, I want to see some kind of privacy setup on a website, I could actually configure it with the software you're building. So, when I set it up with a persona, I know I'll get everything I need (for privacy auditing).}'' P3 added, ``\textit{When the ads are clearly targeted, for example, based on a certain income level (of the persona), it’s definitely something I will pay attention to in the future. I will wonder if this ad is directed at me because of my attributes.}'' \rev{These reflections show how the tool influenced users’ thinking beyond the session, prompting ongoing curiosity and vigilance.}

\rev{Among the 15 participants, eight chose to have each persona variant revisit their selected websites three times, four chose five repetitions, and three opted for two repetitions. This variability shows that some used the tool for deep audits, while others opted for quicker validation—highlighting its flexibility.}

In particular, participants did not rate ``use this system frequently'' highly. In the follow-up interview, they explained two main reasons: first, they believed that frequent and repetitive privacy audits were not necessary for a certain website.  Second, it took a bit too long to conduct each audit using the Privacy Auditing Sandbox, mainly due to the time it takes to sample each website repeatedly to scrape data. Nevertheless, this would not be a barrier for them to audit the target website once. Additionally, some users found the Privacy Auditing Sandbox a bit complex at the beginning, but also mentioned that, after a period of learning, they were able to quickly grasp how to use it, since ``\textit{It breaks everything down by category, and the graphics are pretty clear}'' (P14).



\section{Discussion}



\subsection{The Role of Generated Persona in Usable Privacy and End-User Auditing}
Generated personas play a valuable role in helping users gain generalizable privacy knowledge while minimizing risks. They support ``what-if'' types of exploration, enable large-scale testing, and allow users to compare how algorithmic results may differ between different personas. Well-constructed personas can also evoke empathetic responses, as shown in~\cite{chen2023empathy}, helping users relate the outcomes observed with the personas to potential implications for their own future behavior. As confirmed in Section~\ref{Sec: result rq1}, users were able to load various generated personas into the browser, observe the relationships between ad themes, visual elements, and user characteristics, and validate the impact of these characteristics on recommended content. This demonstrates that users could effectively identify and internalize the consequences of sharing information related to specific privacy attributes by connecting the ads they viewed to the traits of the generated personas.

A key reason users are drawn to using generated personas for privacy audits is that they can test system feedback without exposing their real data. When users share personal data, they lose control over how it is used, making it difficult for them to experiment with different privacy behaviors (e.g., trying to allow access to a particular data to see whether the trade-off in improved service quality is worth it) and observe the outcomes for audits. Using generated personas alleviates concerns about potential privacy breaches, encouraging more active engagement in privacy audits.

Although previous research has pointed out the risk that LLM-generated personas may reinforce existing stereotypes, in the context of our study, these biases may not be detrimental. When recommendation systems process persona data that reflect real-world biases, they generate outcomes representative of real-world services or experiences, as these systems often embed such biases. Consequently, if our generated personas contain biases, they serve as useful tools for auditing recommendations made by external websites and apps, which frequently rely on stereotypes and biases to inform their recommendations.

\subsection{The Complexity in the Design of Privacy Auditing Sandbox}
A common issue reported by users in the study was the system's high level of complexity, particularly at the beginning of use. Although participants found the system to be learnable, there was an initial learning curve in understanding how to navigate the system and interpret the audit results. The complexity of the Privacy Auditing Sandbox arises from several factors, including the interface design, users' data literacy, and unfamiliarity with outputs generated by LLMs.

\subsubsection{Interface Complexity}
As discussed in Section~\ref{Sec: result rq3}, although the system interface was generally clear, some participants initially found it overwhelming due to their unfamiliarity with the concept of end-user auditing. However, after gaining experience with the system, they appreciated its structure and clear visuals, suggesting that the interface becomes more approachable over time. To reduce this initial barrier, a simplified onboarding process or a tutorial could help users become familiar with the layout and functionality of the system more quickly.

\subsubsection{Data literacy}
Users with limited data literacy struggled to interpret the normal distribution charts used to represent privacy risks. Although labels and graphics were provided, additional explanations or visual aids could enhance comprehension for those less familiar with statistical concepts.

\subsubsection{Lack of Familiarity with LLM Generation}
Another challenge was the user's unfamiliarity with LLM generation processes. The Privacy Auditing Sandbox integrates LLM-generated synthetic user data, associated with individual personas, into its auditing system, which some participants found confusing. Providing clearer explanations of how the LLM operates, along with contextual insights into why specific recommendations are made, could help build greater user confidence in the system’s outputs.

\subsection{An End-User Empowerment Approach for Algorithm Accountability} 







The approach introduced by Privacy Auditing Sandbox exemplifies an end-user empowerment approach to enhance algorithm accountability. Instead of relying on the first party, i.e., the owner of the website or third parties such as regulatory bodies or user advocacy groups, to perform audits, our approach allowed everyday web users to perform audits of the relationship between the user's personal attributes and the behaviors of online content recommender systems. This approach has several distinct benefits. 

First, it allows users to validate a ``long tail'' of hypotheses that they are personally interested in. Unlike first- or third-party approaches, which likely will only cover hypotheses relevant to the most common privacy concerns, end users using our tool can now audit for hypotheses they come up with themselves. \revise{Previous research has shown that users harbor notions about how their age, location, or browsing history might influence the ads they encounter~\cite{lee2023and}. However, current auditable AI documentation is designed primarily for regulators and experts rather than end users affected by AI decisions, making it difficult for users to evaluate their privacy hypotheses~\cite{scharowski2023certification}. Through the Privacy Auditing Sandbox, users can use personas to validate their privacy hypotheses without compromising their real personal data. As described in Section~\ref{Sec: result rq1}, users can evaluate whether their data have been used in unexpected ways by linking personal attributes to ad themes and visual elements. Prior research has highlighted ad discrimination in online recommendation systems~\cite{imana2024auditing}. We anticipate that the Privacy Auditing Sandbox can also support verifying these concerns by enabling users to investigate potential discrimination based on specific privacy attributes.}

Second, this approach allows end users to take initiatives by actively participating in algorithmic auditing and experience the varying outcomes of the systems when using different personas, instead of being only passive consumers of auditing reports created by first- or third-parties. This can improve the user's sense of engagement and potentially amplify the positive impact of auditing on user privacy literacy and future behavioral changes based on experiential learning theories~\cite{gentry1990experiential, kolb2014experiential}.  As described in Section~\ref{Sec: result rq2}, participants in our study demonstrated a nuanced awareness of the trade-offs between personalized content and privacy risks. This suggests that empowering users to take more initiative in managing their data and privacy could bridge the gap between perceived benefits and concerns. From the user's perspective, privacy awareness is the prerequisite of privacy literacy. As shown by the discomfort and sense of powerlessness that some participants experienced (Section~\ref{Sec: result rq2}), end users often feel trapped in a cycle of data collection and targeting without clear consent or control. To address this issue, it is crucial to involve users actively in privacy audits, transforming them from passive subjects to engaged participants in the process.

Third, beyond individual users, end-user empowerment has the potential to transform the broader data ecosystem by promoting awareness and accountability among all stakeholders. Privacy awareness and literacy are fundamental to this transformation. Users who better understand data practices are more likely to demand transparency and ethical behavior from companies. Empowered users can more effectively hold platforms accountable by questioning the fairness and appropriateness of algorithms that curate their content and experiences ~\cite{Wieringa2020account}. In this context, end-user empowerment can facilitate a bottom-up approach to privacy activism. When users are educated about the implications of data collection, they are more inclined to engage in collective actions, such as data activism~\cite{vincent2019data,vincent2021data,li2022bottom}, to drive change. These grassroots movements can catalyze broader industry shifts towards more ethical data practices. By actively participating in the dialogue about data use, empowered users contribute to the democratization of algorithmic governance. This bottom-up pressure complements top-down regulatory efforts, creating a more balanced and user-centric approach to data privacy and algorithmic fairness.




\subsection{Ethical and Legal Considerations}
While our approach can help users verify their privacy hypotheses and increase engagement in privacy audits by lowering its barrier to end users, we also identified certain ethical and legal risks associated with its implementation.

One concern is the potential to reinforce user bias. Biased or stereotypical privacy personas may be useful for analyzing the relationship between persona traits and recommended content, but repeated exposure to these personas could unintentionally reinforce users' existing biases. It is important to warn users about the risk of bias and stereotyping in the generated personas, and further research is necessary to mitigate this effect.

Another concern is the potential for malicious misuse, which could lead to cybersecurity risks. Although the method is intended to enhance privacy literacy, the creation of realistic personas could be exploited to develop undetectable bots or facilitate phishing attacks. This underscores the need for stronger technical safeguards and policy measures to address these risks. For instance, incorporating advanced identity verification techniques or restricting access to persona generation tools to authorized users could help prevent such misuse.

\subsection{Limitation and Future Work}
We identify three key limitations of our work and suggest future improvements for them: privacy auditing sandbox prototype, experiment design, and generalizability of auditing tasks.

\subsubsection{Privacy auditing sandbox prototype}
Our Privacy Auditing Sandbox prototype was developed primarily as a proof-of-concept to support the experiments presented in this paper. Consequently, its ability to audit a wide range of privacy attributes is currently limited, with gaps in coverage for commonly used data types, such as search history. From a technical perspective, those data are typically stored and managed on the server-side by service providers rather than locally, making the data replacement approach used in our Sandbox infeasible.

In addition, some data that influence online recommendations, such as inferred behavioral patterns and social connections, are difficult to replicate using only modifiable personal attributes. To overcome these limitations, we plan to enhance the sandbox by expanding its capability to simulate a broader range of privacy data, including more complex attributes that underpin behavioral and social inferences.

\revise{Moreover, the existing proof-of-concept uses the GPT-4 model through its commercial API, which requires users to have access to the model to use the tool. Our future work  will explore how the use of other LLMs, especially open-source ones, to achieve similar performance while making the tool more accessible and easily deployable.}

\subsubsection{Experiment design}
The primary objective of our experimental design was to assess the feasibility of our approach in helping users test their privacy hypotheses and raise their privacy awareness.  Rather than directly measuring users' privacy knowledge gains or behavioral changes resulting from the Privacy Auditing Sandbox, this experiment focused on evaluating the viability of our method. While participant interviews provided qualitative insights suggesting the potential of our approach to foster both privacy knowledge and behavioral changes, further validation is required in long-term, real-world environments. To address this, we plan to conduct longitudinal deployment studies of the Privacy Auditing Sandbox. These studies will allow us to (1) implement pre- and post-tests to evaluate users' privacy knowledge acquisition; (2) conduct longitudinal studies to track changes in users' privacy behaviors over time. This future research will provide a more comprehensive understanding of our approach's effectiveness in promoting privacy literacy and encouraging lasting behavioral changes.

\subsubsection{Generalizability of auditing tasks} 
In our experiment, we focused on online advertisements for downstream tasks as proof-of-concept due to their ubiquity, user familiarity, and sensitivity to changes in privacy data. However, we believe that our approach can be generalized to a variety of other contexts, allowing users to audit the effects of their privacy behaviors in areas such as friend recommendations on social media, dynamic pricing in e-commerce, and personalized content on streaming platforms. In the next phase of developing the Privacy Auditing Sandbox, our goal is to extend its capabilities to cover these additional tasks, facilitating audits of algorithmic accountability. We will also conduct further deployment and testing to evaluate how effectively our approach enhances user engagement in privacy audits and promotes more equitable algorithmic outcomes.


\section{Conclusion}
 We introduced a novel interactive sandbox approach to democratize end-user auditing of online content recommendations, enabling users to test their hypotheses by observing how websites' recommendation algorithms respond to synthetic user personas with variations in selected key attributes. In a case study on targeted ads with 15 participants, the Privacy Auditing Sandbox system demonstrated its usability and effectiveness. Participants successfully verified their privacy-related hypotheses by analyzing ad themes and visual elements, and actively engaging in privacy audits. Our findings offered design implications for empowering users to audit algorithmic accountability through the use of artificially generated personas.

\bibliographystyle{ACM-Reference-Format}
\bibliography{bibliography}

\clearpage
\appendix

\section{Persona used in the User Study}
\label{appendix: persona}

\begin{table*}
\centering
\begin{tabular} {|p{1.4cm}|p{4.3cm}|p{4.3cm}|p{4.3cm}|}
\hline
\textbf{Personal attributes} & \textbf{Persona variant 1} & \textbf{Persona variant 2} & \textbf{Persona variant 3} \\ \hline
\textbf{Age} & 
Young: Michael Johnson is an 22-year-old Black male living in 456 Elm St, Atlanta, GA 30312. He is a college student, currently working as a sales associate with an annual income of \$30,000. & 
Mid-aged: Michael Johnson is a 48-year-old Black male living in 456 Elm St, Atlanta, GA 30312. He holds a bachelor's degree and works as a regional sales director with an annual income of \$90,000. & 
Old: Michael Johnson is a 68-year-old Black male living in 456 Elm St, Atlanta, GA 30312. He has a bachelor's degree and is semi-retired, working part-time as a sales associate with an annual income of \$20,000. \\ \hline

\textbf{Gender} & 
Male: Joshua Williams is a 20-year-old Black male living in 456 Elm St, Atlanta, GA 30312. He has some college experience and works as a sales associate with an annual income of \$30,000. & 
Non-binary: Jordan Williams is a 20-year-old Black non-binary individual living in 456 Elm St, Atlanta, GA 30312. They are pursuing an associate degree in marketing and work part-time as a sales associate with an annual income of approximately \$18,000. & 
Female: Jasmine Williams is a 20-year-old Black female living in 456 Elm St, Atlanta, GA 30312. She is studying business administration at a local college and works weekends as a retail assistant, earning around \$16,000 annually. \\ \hline

\textbf{Location (urbanization)} & 
Urban: Jessica Ramirez is a 40-year-old Hispanic female living in 1202 Congress Ave, Austin, TX 78701. She holds a master’s degree in educational leadership and works as a school administrator at a large public high school, with an annual income of \$95,000. & 
Suburb: Jessica Ramirez is a 40-year-old Hispanic female living in 789 Maple Lane, Round Rock, TX 78664. She holds a master’s degree and works as a school principal in the Round Rock Independent School District, with an annual income of \$105,000.& 
Countryside: Jessica Ramirez is a 40-year-old Hispanic female living in 123 Hilltop Rd, Smithville, TX 78957. She holds a bachelor’s degree in education and works as a K–12 school coordinator at a small rural school, with an annual income of \$58,000. \\ \hline

\textbf{Income level} & 
Low income: Jordan Smith is a 30-year-old Black male living in 456 Elm St, Charlotte, NC 28202. He has a high school diploma and works full-time as a warehouse associate, earning an annual income of \$28,000. & 
Medium income: Jordan Smith is a 30-year-old Black male living in 456 Elm St, Charlotte, NC 28202. He holds a bachelor’s degree in business and works as a sales account executive with an annual income of \$70,000. & 
High income: Jordan Smith is a 30-year-old Black male living in 456 Elm St, Charlotte, NC 28202. He holds an MBA and works as a regional sales director at a tech company, earning an annual income of \$160,000. \\ \hline

\textbf{Education level} & 
Low: Michael Johnson is a 30-year-old Black man living at 782 Eastwood Dr, Atlanta, GA 30312. He holds a high school diploma and completed a CompTIA A+ certification, and works as a full-time IT support technician with an annual income of \$42,000. & 
Medium: Michael Johnson is a 30-year-old Black man living at 456 Elm St, Atlanta, GA 30312. He holds a bachelor's degree in Computer Science and works as a software engineer at a mid-sized fintech company, with an annual income of \$95,000. & 
High: Michael Johnson is a 30-year-old Black man living at 1200 Techwood Dr NW, Atlanta, GA 30318. He holds a Ph.D. in Computer Science and works as a machine learning researcher at a leading tech firm, with an annual income of \$180,000. \\ \hline
\end{tabular}
\caption{Persona Variants Based on Personal Attributes}
\end{table*}
\section{Codebook}
\label{appendix: codebook}

\begin{enumerate}
    \item Reason for having relationship
    \begin{enumerate}
        \item Ad visual elements
        \begin{enumerate}
            \item Users tend to see the similar age group people in ads
            \item Ads for men often use red and blue color
            \item Ads for women often use pink or purple color
            \item Ads for women often use images of women themselves
        \end{enumerate}
        \item Ad topic
        \begin{enumerate}
            \item Middle ages are more business oriented
            \item Workout ads are more related with young people
            \item Beauty ads are more related with young people
            \item Products for students are more related with young people
            \item Video game is more related with young people
            \item Older adults are more likely to see disaster prevention ads
            \item Middle-aged/old adults are more likely to see skin repair ads
            \item Older adults are more likely to see financial aid ads
            \item Older adults are more likely to see health related ads
            \item Women are more likely to see female shoe ads
            \item Women see female fitness ads
            \item Men are more likely to see tech-related ads
            \item Drive up service target suburb/countryside people
            \item Expensive product ads target high-income people
            \item Coupon and discount target people with low income and education level
            \item Lower income people get more ads about junk food
            \item Ads for financial aid target low income people
            \item Ads for continuing studies target low education people
            \item Ads for traveling may target medium or high income people
        \end{enumerate}
    \end{enumerate}

    \item Attitudes, awareness, and sense of empowerment
    \begin{enumerate}
        \item Positive attitudes
        \begin{enumerate}
            \item The content is just enough to verify their hypothesis
            \item Personal data tracking can improve the recommendation quality
            \item Support verifying relation between privacy attributes and ads
            \item Improve privacy awareness
        \end{enumerate}
        \item Negative attitudes
        \begin{enumerate}
            \item Privacy Resignation
            \item Concern over privacy invasion
        \end{enumerate}
    \end{enumerate}
\end{enumerate}

\end{document}